\documentstyle[12pt,epsfig,ulem,graphicx, amssymb,color]{article}

\newcommand{\hsc}{\hspace*{-10pt}}

\title{Behavioral Portfolio Selection in Continuous Time\footnote{This
paper
has benefited from comments of participants at
the Quantitative Methods in Finance 2005 Conference in Sydney,
the 2005 International Workshop on Financial Engineering and Risk Management
in Beijing, the 2006 International Symposium on Stochastic Processes and
Applications to Mathematical Finance in Ritsumeikan University, and
the 2006 Workshop on Mathematical Finance and Insurance in Lijiang;
and from the comments of Knut Aase, Andrew Cairns, Mark Davis, Peter Imkeller,
Jacek Krawczyk, Terry Lyons, James Mirrlees, Eckhard Platen,
Sheldon Ross, Larry Samuelson, Martin Schweizer, John van der Hoek, and
David Yao.
The authors thank especially Jia-an Yan for remarks and discussions
over the years on the general topic considered in this paper.
Two anonymous referees have given constructive comments leading to
a much improved version.
All errors are the responsibility of the authors.
Zhou gratefully acknowledges financial support from the RGC Earmarked Grants
CUHK4175/03E and CUHK418605, and the Croucher Senior Research Fellowship.}}

\author{Hanqing Jin\thanks{Department of Mathematics, National University of Singapore,
Singapore. Email: $<$matjinh@nus.edu.sg$>$.}
\ \ and \ \ Xun Yu Zhou\thanks{Mathematical Institute, 
University of Oxford, 
24-29 St Giles', Oxford,  OX1 3LB, UK, and 
Department of Systems Engineering and Engineering
Management, The Chinese University of Hong Kong, Shatin, Hong Kong.
email: $<$zhouxy@maths.ox.ac.uk$>$. 
}}

\newcommand{\dpm}{{\prime\hspace{-0.03cm}\prime}}

\newtheorem{theo}{\sc Theorem}[section]
\newtheorem{lemma}{\sc Lemma}[section]
\newtheorem{prop}{\sc Proposition}[section]

\newtheorem{assump}{\sc Assumption}[section]
\newtheorem{remark}{\sc Remark}[section]

\newtheorem{exam}{\sc Example}[section]
\newcommand{\eof}{\hfill {\it Q.E.D.} \vspace*{0.3cm}}

\newcommand{\pf}{{\it Proof: }}

\newcommand{\cF}{{\mathcal F}}
\newcommand{\R}{{\hbox{I{\kern -0.22em}R}}}
\newcommand{\sR}{\rm I{\kern -0.22em}R}
\newcommand{\id}{{\mathbf 1}}
\newcommand{\as}{\mbox{{\rm a.s.}}}
\newcommand{\aee}{\mbox{{\rm a.e.}}}

\begin{document}
\date{\today}
\maketitle

\begin{abstract}

This paper formulates and studies a general continuous-time
behavioral portfolio selection model under Kahneman and Tversky's
(cumulative) prospect theory, featuring S-shaped utility (value) functions
and probability distortions. Unlike the conventional
expected utility maximization model, such a behavioral model could be easily
mis-formulated (a.k.a. ill-posed) if its different components
do not coordinate well with each other. Certain classes of an
ill-posed model are
identified.
A  systematic approach, which is
fundamentally different from the ones employed for the utility model,
is developed to solve a well-posed model, assuming a complete market and
general It\^o processes for asset prices.
The optimal terminal wealth positions, derived in fairly explicit forms,
possess surprisingly simple structure reminiscent of a gambling policy betting on
a good state of the world while accepting a fixed, known loss in case of a bad one.
An example
with a two-piece CRRA utility is presented to illustrate the
general results obtained, and is solved completely for all admissible
parameters. The effect of the behavioral criterion
on the risky allocations is finally discussed.

\medskip


\end{abstract}

\section{Introduction}

Mean--variance and expected utility maximization are by far
the two predominant
investment decision rules in financial portfolio selection.
Portfolio
theory in the dynamic setting (both discrete time and continuous time)
has been established in the past twenty years, again
centering around these two
frameworks while employing heavily among others the martingale theory, convex duality and
stochastic control; see Duffie (1996), Karatzas and Shreve (1998), and F\"ollmer and Schied (2002)
for systematic accounts on dynamic utility maximization, and
Li and Ng (2000), Zhou and Li (2000), and Jin, Yan and Zhou (2004) for
recent studies on the mean--variance (including extensions to mean--risk)
counterpart.

Expected utility theory (EUT), developed by
von Neumann and Morgenstern (1944) based on an axiomatic system, has
an underlying assumption that decision makers are rational and risk averse
when facing uncertainties.
In the context of asset allocations, its basic tenets are:
Investors evaluate wealth according to final asset positions;
they are uniformly risk averse; and
they are able to objectively evaluate probabilities.
These, however, have long been criticized to be
inconsistent with the way people do decision making in the real world.
Substantial experimental evidences have suggested a
systematic violation of the EUT principles.
Specifically, the following anomalies (as opposed to the assumed rationality in
EUT) in human behaviors are evident from daily life:
\begin{itemize}
\item People evaluate assets on gains and losses (which are defined
with respect to a reference point), not
on final wealth positions; 
\item People are {\it not} uniformly risk averse: they are risk-averse on gains and risk-taking on losses, and significantly more sensitive to losses than to gains; 
\item People overweight small probabilities and underweight
large probabilities. 
\end{itemize}

In addition, there are widely known paradoxes and puzzles that
EUT fails to explain, including
the Allais paradox [Allais (1953)], Ellesberg paradox
[Ellesberg (1961)], Friedman and Savage puzzle [Friedman and Savage (1948)], and the
equity premium puzzle [Mehra and Prescott (1985)].

Considerable attempts and efforts
have been made to address the drawback of EUT, among them
notably the so-called non-additive utility theory [see, for example,
Fishburn (1998)].
Unfortunately, most of these theories
are far too complicated to be analyzable and applicable, and some of them even lead to new paradoxes.
In 1970s, Kahneman and Tversky (1979)
proposed the {\it prospect theory} (PT) for decision making under
uncertainty,
incorporating human emotions and psychology into their theory.
Later, Tversky and Kahneman (1992) fine tuned the
PT to the {\it cumulated prospect theory} (CPT)
in order to be consistent with the first-order stochastic dominance.
Among many other ingredients, the key elements of Kahneman and Tversky's
Nobel-prize-winning theory are
\begin{itemize}
\item A {\it reference point} (or {\it neutral outcome/benchmark/breakeven point/status quo})
in wealth that defines {\it gains} and {\it losses};
\item A value function (which replaces the notion of utility function), {\it concave for gains} and {\it convex for losses}, and
steeper for losses than for gains (a behavior called {\it loss aversion});
\item A {\it probability distortion} that is a {\it nonlinear} transformation
of
the probability scale, which enlarges a small probability
  and diminishes a large probability.
\end{itemize}

There have been burgeoning research interests
in incorporating the PT into portfolio choice;
nonetheless they have been hitherto
overwhelmingly limited to the single-period setting; see
for example 
Benartzi and Thaler (1995), Shefrin and Statman (2000),
Levy and Levy (2004), Bassett {\it et al.} (2004), Gomes (2005), and
De Giorgi and Post (2005), with emphases on qualitative properties and
empirical experiments. {\it Analytical} research on {\it dynamic}, especially continuous-time,
asset allocation featuring behavioral criteria is literally nil according
to our best knowledge. [In this connection the only paper we know of
that has some bearing on the PT for the continuous time setting
is Berkelaar, Kouwenberg and Post
(2004) where a
very specific two-piece power utility function is considered; however, the probability
distortion, which is one of the major ingredients of the PT
and which causes the main difficulty, is absent in that paper.]
Such a lack of study on continuous-time behavioral portfolio
selection is certainly not because the problem is uninteresting or unimportant; rather it
is because, we believe, that the problem is {\it massively} difficult as
compared with the conventional expected utility maximization model.
Many conventional and convenient approaches, such as convex optimization,
dynamic programming,
and stochastic control, fall completely apart in handling
such a behavioral model:
First, the utility function (or {\it value function} as called in the PT)
is partly concave and partly convex (also referred to as
an {\it S-shaped} function), whereas the global convexity/concavity
is a necessity in traditional optimization. Second,
the nonlinear distortion in probabilities abolishes virtually all the
nice properties associated with the normal additive probability and linear expectation.
In particular, the dynamic consistency of the conditional expectation
with respect to a filtration, which is the foundation of the
dynamic programming principle, is absent due to the distorted probability.
Worse still, the coupling of these two ill-behaved features greatly
amplifies the difficulty of the problem\footnote{In Berkelaar, Kouwenberg and Post
(2004) an essentially  convexification technique is employed to deal with the
non-convexity of the problem. However, it does not work any longer in the presence of
a distorted probability.}.
Even the well-posedness of the problem\footnote{A
maximization problem is called {\it well-posed} if its supremum is finite; otherwise it is
{\it ill-posed}. An ill-posed problem is a mis-formulated one:
the trade-off is not set right so that one can always push the
objective value to be arbitrarily high.} is no longer something that can be taken for granted.

This paper first establishes a general continuous-time
portfolio selection model under the CPT, involving behavioral
criteria defined on possibly continuous random variables.
The probability distortions lead to the involvement of the
Choquet integrals [Choquet
(1953/54)], instead of the
conventional expectation. We
then
carry out, {\it analytically}, extensive investigations on
the model while developing new approaches
in deriving the optimal solutions. First of all, by assuming
that the market is complete,
the asset prices follow general It\^o processes, and the
individual behavior of the investor
in question will not affect the market,
we
need only to consider an optimization problem in terms of
the terminal wealth.
This is the usual trick employed in the conventional utility maximization, which also
enables us to get around the inapplicability of the dynamic programming
in the current setting.
Having said this, our main endeavor is to find the optimal terminal wealth,
which is a fundamentally different and difficult problem due to the
behavioral criterion.
As mentioned earlier
such a behavioral
model could be easily ill-posed and, therefore,  we first identify several
general cases where the model is indeed ill-posed.
Then we move on to finding optimal solutions for a well-posed model.
In doing so we decompose the original problem into two sub-problems:
one takes care of the gain part and the other the loss part, both parameterized
by an initial budget that is the price of the gain
part (i.e., the positive change)
of the terminal payoff over the reference wealth position
and an event when the terminal payoff represents a gain.
At the outset the gain part problem is a constrained
{\it non-concave} maximization problem
due to the probability distortion; yet
by changing the decision variable
and taking a series of
transformations, we turn it into
a concave maximization problem
where the Lagrange method is applicable. The loss part problem,
nevertheless, is more subtle because it is to {\it minimize} a {\it concave}
functional even after the similar transformations.
We are able to characterize explicitly its solutions to be certain
``corner points'' via delicate analysis.
There is yet one more twist in deriving the optimal solution
to the original model given the
solutions to the above two problems: one needs to find the
``best'' parameters -- the initial budget and the event
of a terminal gain --
by solving another constrained optimization problem.

As mathematically
complicated and sophisticated the solution procedure turns out to be,
the final solutions
are surprisingly and beautifully simple: the optimal terminal wealth
resembles the payoff
of a portfolio of two {\it binary} (or {\it digital}) {\it options} written
on a mutual fund (induced by the state pricing density), characterized by a
{single} number. This number, in turn, can be identified by
solving a
very simple two-dimensional mathematical programming problem.
The optimal strategy is therefore a gambling
policy, betting on good states of the market, by buying
a contingent claim and selling another.
We present an example with the same value function taken by
Tversky and Kahneman (1992), and demonstrate that our general results lead to
a complete solution of the model for all admissible parameters
involved. Furthermore, for the case when the market parameters are
constants, we are able to derive the optimal portfolio in closed form, thereby
understand how the behavioral criteria may change the risky asset allocations.

To summarize, the main contributions of this paper are: 1) we establish,
for the first time, a {\it bona fide} continuous-time behavioral
portfolio selection model {\it \`a la} cumulative prospect theory,
featuring very general S-shaped utility functions and probability
distortions;
2) we demonstrate that the well-posedness becomes
an eminent issue for the behavioral
model, and identify several ill-posed
problems;
3) we develop an approach, fundamentally different from the existing ones for
the expected utility model, to overcome the immense
difficulties arising from the analytically ill-behaved
utility functions and probability
distortions. Some of the sub-problems solvable by this approach, such as
constrained maximization and minimization of Choquet integrals, are interesting, in both
theory and applications, in their
own rights; and 4) we obtain fairly explicit solutions to  a general model, and
closed-form solutions for an important  special case, based on which we are able to
examine how the allocations to equity are influenced by behavioral criteria.

The rest of the paper is organized as follow. In Section 2
the behavioral model is formulated, and its possible ill-posedness
is addressed in Section 3. The main results of the paper are stated in
Section 4. The procedure of analytically solving
the general model is developed
in Sections 5 -- 7, leading to a proof of the main results in Section 8.  A special case with a two-piece CRRA utility function
is presented in Section 9 to demonstrate the general results obtained.
Section 10 addresses the issue of how the behavioral criterion would affect
the risky allocations. Some concluding remarks are given in Section 11.
Finally,
technical preliminaries are relegated to an appendix.

\section{The Model}
In this paper $T$ is a fixed terminal time and $(\Omega,\cF,P,\{\cF_t\}_{t\geq0})$ is a
fixed filtered complete probability space on which is defined a standard $\cF_t$-adapted
$m$-dimensional Brownian motion $W(t)\equiv(W^1(t),\cdots,W^m(t))'$ with $W(0)=0$.
It is assumed that $\cF_t=\sigma\{W(s):0\le s\le t\}$, augmented by all the
null sets.
Here and throughout the paper $A'$ denotes the transpose of a matrix
$A$.

We define a continuous-time financial market following Karatzas and Shreve (1998).
In the market there are $m+1$ assets being
traded continuously. One of the assets
is a bank account whose {price process} $S_0(t)$ is subject to the following
equation:
\begin{equation}\label{bond}
dS_0(t)=r(t)S_0(t)dt,\;\; t\in[0,T]; \;\; S_0(0)=s_0>0,
\end{equation}
where the {interest rate} $r(\cdot)$ is
an $\cF_t$-progressively measurable,
scalar-valued stochastic process with $\int_0^T|r(s)|ds<+\infty, \;\as$.
The other $m$ assets are {stocks}
whose price processes $S_i(t)$, $i=1,\cdots, m$, satisfy the following stochastic
differential equation (SDE):
\begin{equation}\label{security}
dS_i(t)=S_i(t)\big[b_i(t)dt+\sum_{j=1}^m\sigma_{ij}(t)dW^j(t)\big],\;\;
t\in[0,T];\;\;
S_i(0)=s_i>0,
\end{equation}
where $b_i(\cdot)$ and $\sigma_{ij}(\cdot)$, the {appreciation} and {dispersion} (or
{volatility}) rates, respectively, are scalar-valued, $\cF_t$-progressively measurable
stochastic processes with $\int_0^T[\sum_{i=1}^m|b_i(t)|+\sum_{i,j=1}^m|\sigma_{ij}(t)|^2]dt<+\infty,\,\as$.

Set the excess rate of return vector process
$$B(t):=(b_1(t)-r(t),\cdots,b_m(t)-r(t))',$$
and define the {volatility matrix process}
$\sigma(t)
:=(\sigma_{ij}(t))_{m\times m}$.
Basic assumptions imposed on the market parameters throughout this paper
are summarized as follows:

\begin{assump}\label{no-arb}
{\rm \begin{itemize}
\item []
  \item [(i)] There exists $c\in\R$ such that $\int_0^Tr(s)ds\geq c$, \as.
\item [(ii)] $\mbox{{\rm Rank}}\;(\sigma(t))=m$, \aee $t\in[0,T]$, \as.
  \item [(ii)] There exists an $\R^m$-valued,  uniformly bounded,
$\cF_t$-progressively measurable process $\theta(\cdot)$ such that
$\sigma(t)\theta(t)=B(t)$, \aee $t\in[0,T]$, \as.
\end{itemize}
}
\end{assump}

It is well known that under these assumptions
there exists a unique risk-neutral (martingale)
probability measure $Q$ defined by
$\frac{dQ}{dP}\Big|_{\cF_t}=\rho(t)$, where
\begin{equation}\label{rho}
\rho(t):=\exp\left\{-\int_0^t\left[r(s)+\frac{1}{2}|\theta(s)|^2\right]ds-\int_0^t\theta(s)'dW(s)\right\}
\end{equation}
is the pricing kernel or state density price.
Denote $\rho:=\rho(T)$. It is clear that
$0<\rho<+\infty$ \as,  and $0<E\rho<+\infty$.

\medskip

A random variable $\xi$ is said to {\it have no atom} if
$P\{\xi=a\}=0\;\forall a\in\R$.
The following assumption is in force throughout this paper.
\begin{assump}\label{no-atom}
{\rm $\rho$ admits no atom.}
\end{assump}

The preceding assumption is not essential, and is imposed
to avoid undue technicality.
In particular, it is satisfied when
$r(\cdot)$ and
$\theta(\cdot)$ are deterministic with $\int_0^T|\theta(t)|^2dt\neq0$ (in which case $\rho$ is a nondegenerate lognormal random variable). We are also going
to use the following notation:
\begin{equation}\label{rhobar}
\begin{array}{l}
\bar\rho\equiv {\rm esssup}\; \rho:=\sup\left\{a\in\R: P\{\rho>a\}>0\right\},\\
\underline{\rho}\equiv {\rm essinf}\;\rho:=\inf\left\{a\in\R: P\{\rho<a\}>0\right\}.
\end{array}
\end{equation}

\medskip

Consider an agent, with an initial endowment
 $x_0\in\R$ (fixed throughout this paper)\footnote{Precisely speaking,
$x_0$ should be the difference between the agent's initial wealth and
a (discounted) reference wealth; for details see Remarks \ref{rem1} and \ref{rem2} below.}, whose total wealth at time $t\ge0$ is denoted by $x(t)$. Assume that the
trading of shares takes place continuously {in a self-financing fashion} (i.e., there is no
consumption or income) and there are no transaction costs. Then $x(\cdot)$ satisfies
[see, e.g., Karatzas and Shreve (1998)]
\begin{equation}\label{system}
dx(t)=[r(t)x(t)+B'(t)\pi(t)]dt+\pi(t)'\sigma(t)dW(t),\;\;t\in[0,T];\;\;
x(0)=x_0,
\end{equation}
where $\pi(\cdot)\equiv (\pi_1(\cdot),\cdots,\pi_m(\cdot))'$ is the {\it portfolio} of the agent
with $\pi_i (t),\;\; i=1,2\cdots,m,$ denoting the total market value of the agent's wealth in the $i$-th
asset at time $t$.
{A portfolio $\pi(\cdot )$ is said to be {\it admissible} if
it is an $\R^m$-valued,
$\cF_t$-progressively measurable process with
\begin{eqnarray*}
\int_0^T|\sigma(t)'\pi(t)|^2dt<+\infty  \;\mbox{ and }
\; \int_0^T|B(t)'\pi(t)|dt<+\infty,\;\;\as.
\end{eqnarray*}}
An admissible portfolio $\pi(\cdot )$ is said to be {\it tame} if the
corresponding discounted wealth process, $S_0(t)^{-1}x(t)$,
is almost surely bounded from below (the bound may depend on
$\pi(\cdot )$).

The following result follows from
Karatzas and Shreve (1998, p. 24, Theorem 6.6) noting
Karatzas and Shreve (1998, p. 21, Definition 6.1) or Cox and Huang (1989).

\begin{prop}\label{market}
For any $\cF_T$-measurable random variable $\xi$ such that
$\xi$ is almost surely bounded from below and
$E[\rho\xi]=x_0$, there exists a tame admissible portfolio
$\pi(\cdot)$ such that the corresponding wealth process $x(\cdot)$
satisfies $x(T)=\xi$.
\end{prop}

In the conventional portfolio theory,
an investor's preference is modelled by
the expected utility of the terminal wealth.
In this paper, we study a portfolio model featuring human behaviors by
working within the CPT framework of Tversky and Kahneman (1992).
First of all, in CPT there is a natural outcome or benchmark,
assumed to be 0 (evaluated at the terminal time, $T$) in this paper
without loss of generality 
(see Remark \ref{rem1} below for elaborations on this point),
 which serves
as a base point to distinguish gains from losses.
Next,
we are given two utility functions $u_+(\cdot)$ and $u_-(\cdot)$,
both mapping from
$\R^+$ to $\R^+$,
that measure the gains and losses respectively.
There are two additional
functions $T_+(\cdot)$ and $T_-(\cdot)$
from $[0,1]$
to $[0,1]$, representing the distortions in probability for the gains and
losses respectively. The technical assumptions on these functions, which
will be imposed throughout this paper, are summarized as follows.

\begin{assump}\label{uassump}
{\rm $u_+(\cdot)$ and $u_-(\cdot)$: $\R^+\mapsto \R^+$,
are strictly increasing, concave, with $u_+(0)=u_-(0)=0$. Moreover,
$u_+(\cdot)$ is {strictly concave and} twice differentiable,
with the Inada conditions $u_+'(0+)=+\infty$ and
$u_+'(+\infty)=0$.}
\end{assump}

\begin{assump}\label{Tassump}
{\rm $T_+(\cdot)$ and $T_-(\cdot)$: $[0,1]\mapsto [0,1]$, are
{differentiable and} strictly increasing,
with $T_+(0)=T_-(0)=0$ and $T_+(1)=T_-(1)=1$.
}
\end{assump}

Now, given a contingent claim (a random variable) $X$,
we assign it a value $V(X)$
by
\[V(X)=V_+(X^+)-V_-(X^-)\]
where
\[V_+(Y):=\int_0^{+\infty}T_+(P\{u_+(Y)>y\})dy,\;\;\;\;
V_-(Y):=\int_0^{+\infty}T_-(P\{u_-(Y)>y\})dy\]
for any random variable $Y\ge 0,\;\as$.
(Throughout this paper $a^+$ and $a^-$ denote respectively the positive and negative parts of a
real number $a$.)
It is evident that both $V_+$ and $V_-$ are non-decreasing in the sense that $V_\pm(X)\geq V_\pm(Y)$
for any random variables $X$ and $Y$ with $X\geq Y$ \as.
Moreover, $V_+(x)=u_+(x)$ and $V_-(x)=u_-(x)$ $\forall x\in \R^+$.
Finally, $V$ is also non-decreasing.

{\rm If $T_+(x)=x$ (there is no distortion) then $V_+(Y)=E[u_+(Y)]$ (likewise
with $V_-$); hence $V_+$ is a generalization of the expected utility. Yet
this generalization poses a fundamentally different (and difficult)
feature, namely, the set function $T_+\circ P$ is a {\it capacity} [Choquet
(1953/54)] which is a non-additive measure as opposed to the standard notion
of probability. So the definition of $V_+$ involves
the so-called Choquet integral [see
Denneberg (1994) for a comprehensive account on Choquet integrals].}
Notice that with the Choquet integral the dynamic consistency
of conditional expectation, which is
the base for the dynamic programming principle, is lost\footnote{The
{\it dynamic consistency} refers to the following equality:
$E\left(E(X|{\cal F}_t)|{\cal F}_s\right)=E(X|{\cal F}_s)$ if ${\cal F}_s\subseteq {\cal F}_t$. The problem of generalizing conditional
expectation to Choquet integral remains largely open, not to mention the
validity of the corresponding dynamic consistency in any sense; see
Denneberg (1994, Chapter 12).}.

{\rm In CPT the utility (or value)
function $u(\cdot)$ is given on the whole real line,
which is convex on $\R^-$ and concave on $\R^+$ (corresponding to the
observation that people tend to be risk-averse on gains and risk-seeking on
losses). Such a function is said to be of {\it S-shaped}.
In our model, we separate the utility on gains and losses by letting
$u_+(x):=u(x)$ and $u_-(x)=-u(-x)$ whenever $x\geq0$. Thus our model
is equivalent to the one with an overall S-shaped utility function.}



{\rm In our model, the value $V(X)$ is defined on a general random variable
$X$,
possibly a {\it continuous} one, which is necessary for the continuous-time
portfolio selection model, as opposed to Tversky and Kahneman (1992)
where only {\it discrete} random variables are treated. Moreover,
our definition of $V$ agrees with that in Tversky and Kahneman (1992) if
$X$ is discrete [see Tversky and Kahneman (1992, pp. 300-301)].}

Under this CPT framework, our portfolio selection problem is to find the most preferable portfolios, in terms of
maximizing the value $V(x(T))$, by continuously managing
the portfolio.
The mathematical formulation is as follows:
\begin{equation}\label{btps0}
\begin{array}{ll}
\mbox{\rm Maximize} & V(x(T))\\
\mbox{\rm subject to} & 
               (x(\cdot), \pi(\cdot)) \mbox{ satisfies (\ref{system})},\;\;
               \pi(\cdot) \; \mbox{ is admissible and tame. }
\end{array}
\end{equation}

In view of Proposition \ref{market}, in order to solve (\ref{btps0})
one needs only first to solve the following optimization problem in
the terminal wealth, $X$:
\begin{equation}\label{btps}
\begin{array}{ll}
\mbox{\rm Maximize} & V(X)\\
\mbox{\rm subject to} & E[\rho X]=x_0,\;\;
 X \mbox{ is an \as \  lower bounded, ${\cF}_T$-random variable}.
\end{array}
\end{equation}


Once (\ref{btps}) is solved with a solution $X^*$, the optimal
portfolio is then
the one replicating $X^*$ (as determined by Proposition \ref{market}).
Therefore, in the rest of the paper
we will focus on Problem (\ref{btps}).
Recall that a maximization problem is called {\it well-posed}
if the supremum of its objective is finite; otherwise it is called
{\it ill-posed}. One tries to find an optimal solution only if the
problem is known {\it a priori} to be well-posed.

\begin{remark}\label{rem1}
{\rm If the reference point at $T$ is a general
${\cal F}_T$-measurable random variable
$\xi$ (instead of 0), then, since the market is complete, we can replicate $\xi$ by
a replicating portfolio $\bar\pi(\cdot)$ with the corresponding wealth
process $\bar x(\cdot)$. [Incidentally,
one can also take this case as one where there is a dynamically and stochastically
changing reference trajectory $\bar x(\cdot)$.]
In this case, by considering
$x(t)-\bar x(t)$ as the state variable the problem (\ref{btps0}) is reduced to
one with the reference point being 0.
[In view of this,
the process $x(\cdot)$ determined by (\ref{system}) actually
represents the magnitude of the change in wealth from the
the price process of the terminal reference point.
In particular, this is also why the given initial state $x_0$ in (\ref{system})
can be
any {\it real} number.]}
\end{remark}

\begin{remark}\label{rem2}
{\rm Following the discussion of Remark \ref{rem1}, we see that
our model models the situation where the investor concerns a reference wealth
only at the terminal of the planning horizon (or, equivalently, an
exogenously given dynamic reference trajectory). Examples of such a
situation are when a person is to make a down payment of a house in three
months (in which case the reference point is a deterministic constant), or
when an investor is to cover the short position in a stock in one month (where
the reference point is a random variable). It is certainly plausible
that an investor will update his reference point dynamically. If the
updating rule is known a priori, such as in Berkelaar {\it et al.}
(2004), then it is possible to turn the problem into one covered by
(\ref{btps0}) by appropriately modifying some parameters. If, however,
updating the reference point is {\it part} of the overall decision, then
it would lead to a completely different and interesting model, which is
open for further study.}
\end{remark}

\begin{remark}\label{rem3}
{\rm We implicitly assume in our model that the agent is a ``small
investor''; so his behavior only affects {his} utility function -- and hence
{\it his} asset allocation -- but not the overall market. This is why
the budget constraint in (\ref{btps}), $E[\rho X]=x_0$, is still evaluated
in the conventional sense (no probability distortion). In other words,
$E[\rho X]=x_0$ is the pricing rule of the market, which is (assumed to be)
not influenced by the small investor under consideration. }
\end{remark}

Before we conclude this section, we recall the following definition.
For any non-decreasing function
$f$: $\R^+\mapsto \R^+$, we
define its inverse function
\begin{equation}\label{inverse}
f^{-1}(x):=\inf\{y\in\R^+: f(y)\geq x\},\;\; x\in\R^+.
\end{equation}
It is immediate that $f^{-1}$ is non-decreasing and continuous on the left, and
it holds always that
\[
 f^{-1}(f(y))\leq y.\]

\section{Ill-Posedness}

In general ill-posedness of an optimization problem
signifies that the trade-off therein is not set right, leading to
a wrong model.
Well-posedness is an important issue from the modeling point of
view.
In classical portfolio
selection literature [see, e.g., Karatzas and Shreve (1998)]
 the utility function is  typically assumed to be
globally concave along with other nice properties; thus
the problem is guaranteed to be well-posed in most cases\footnote{Even with a global concave  utility function the underlying problem could still be ill-posed; see counter-examples and discussions in Korn and Kraft (2004) and
Jin, Xu and Zhou (2007).}.
We now demonstrate that for the behavioral model (\ref{btps0})
or (\ref{btps})
the well-posedness becomes a more significant issue, and that
probability distortions
in gains and losses play prominent, yet somewhat opposite, roles.

\begin{theo}\label{ip-1}
  Problem (\ref{btps}) is ill-posed if there exists a nonnegative
$\cF_T$-measurable random variable $X$ such that
  $E[\rho X]<+\infty$ and $V_+(X)=+\infty$.
\end{theo}

\ \ {\sc Proof:} Define $Y:=X-c$ with $c:=(E[\rho X]-x_0)/E\rho$. Then $Y$ is feasible for Problem (\ref{btps}).
If $c\le 0$, then obviously $V(Y)=V_+(Y)\ge V_+(X)=+\infty$.
If $c>0$, then
\begin{eqnarray*}
  V(Y)&=&V_+(Y^+)-V_-(Y^-)\\
  &\ge&V_+(Y^+)-V_-(c)\\
  &=&\int_0^{+\infty}T_+\left(P\{u_+((X-c)^+)>y\}\right)dy-u_-(c)\\
&=&\int_0^{+\infty}T_+\left(P\{u_+(X-c)>y,X\geq c\}\right)dy-u_-(c)\\
  &\ge&\int_0^{+\infty}T_+\left(P\{u_+(X)>y+u_+(c)\}\right)dy-u_-(c)\\
  &\ge&\int_{u_+(c)}^{+\infty}T_+(P\{u_+(X)>y\})dy-u_-(c)\\
  &=&+\infty,
\end{eqnarray*}
where we have used the fact that $u_+(x+y)\leq u_+(x)+u_+(y)\;\forall x,y\in\R^+$ due to the concavity of $u_+(\cdot)$ along with $u_+(0)=0$. The proof is
complete.
\eof

This theorem says that
the model is ill-posed if one can find a nonnegative claim having a finite 
price yet an infinite prospective value.
In this case the agent
can purchase such a claim initially (by taking out a loan if necessary) and
reach the infinite value at the end.
The following example shows that such an almost ``unbelievable'' claim could indeed exist even with very ``nice''
parameters involved, so long as the probability on gains is distorted. 
\begin{exam}\label{ip-1ex}
{\rm 
  Let $\rho$ be such that its (probability) distribution function, $F(\cdot)$,
is continuous and strictly increasing, with $E\rho^3<+\infty$ (e.g., when
$\rho$ is lognormal).
Take $T_+(t):=t^{1/4}$ on $[0, 1/2]$ and $u_+(x):={x}^{1/2}$.
Set
$Z:=F(\rho)$. Then it is known that $Z\sim U(0,1)$, the uniform distribution
on $(0,1)$.
Define $X:=Z^{-1/2}-1$.
Then $X\ge 0$, $P(X>x)=(1+x)^{-2}$ for $x\ge 0$, and
\[ E[\rho X]=E[\rho Z^{-1/2}]-E\rho\le (EZ^{-3/4})^{2/3}(E\rho^3)^{1/3}-E\rho=4^{2/3}(E\rho^3)^{1/3}-E\rho<+\infty.\]
However,
\[
\begin{array}{rl}
    V_+(X)\geq &\int_2^{+\infty}T_+(P\{X>y^2\})dy
=\int_2^{+\infty}T_+\left((1+y^2)^{-2}\right)dy\\
=&\int_2^{+\infty}(1+y^2)^{-1/2}dy
>\int_2^{+\infty}(2y^2)^{-1/2}dy=+\infty.
\end{array}
\]
}
\end{exam}

\medskip

In this example $u_+(x)=x^{1/2}$ is a perfectly ``nice'' utility function satisfying every
condition required for well-posedness (as well as
solvability) of the classical utility model;
yet the distortion $T_+(\cdot)$ ruins everything and
turns the problem into an ill-posed one.

To exclude the ill-posed case identified by Theorem \ref{ip-1},
we need the following assumption throughout this paper:
\begin{assump}\label{ill}
{\rm $V_+(X)<+\infty$ for any nonnegative, ${\cF}_T$-measurable random variable $X$ satisfying
$E[\rho X]<+\infty$.}
\end{assump}


{Assumption \ref{ill} is not sufficient to completely
rule out the ill-posedness.
The following theorem specifies another class of ill-posed problems.
}

\begin{theo}\label{ip-2}
If $u_+(+\infty)=+\infty$, $\bar\rho =+\infty$, 
and  $T_-(x)=x$,
then Problem (\ref{btps}) is ill-posed.
\end{theo}

{\sc Proof:} Fix {any $a>\underline \rho$ and define} $X:=c\id_{\rho<{a}}$ with
$c:=\frac{x_0^++1}{E[\rho\id_{\rho<{a}}]}>0$.
Then for any $n>0$,
\begin{equation}\label{vxn0}
\begin{array}{rl}
V_+(nX)=&\int_0^{+\infty}T_+(P\{u_+(nX)>y\})dy\\
=&\int_0^{u_+(nc)}T_+(P\{u_+(nc\id_{\rho<{a}})>y\})dy\\
=&u_+(nc)T_+(P\{\rho<{a}\})\rightarrow +\infty\mbox{ as }n\rightarrow +\infty.
\end{array}
\end{equation}
Next,
for any $n>1$, define
$X_n:=c_n\id_{\rho>n^2}$, where $c_n:=\frac{nE[\rho X]-x_0}{E[\rho\id_{\rho>n^2}]}$.
(Here $E[\rho\id_{\rho>n^2}]>0$ thanks to $\bar\rho =+\infty$.) Obviously,
$c_n^+P\{\rho>n^2\}=\frac{(nE[\rho X]-x_0)^+}{E[\rho|\rho>n^2]}\le \frac{|nE[\rho X]-x_0|}{n^2} \rightarrow 0$
as $n\rightarrow +\infty$. Hence
\begin{equation}\label{vxn}
V_-(c_n^+\id_{\rho>n^2})=u_-(c_n^+)P\{\rho>n^2\}\le u_-(c_n^+P\{\rho>n^2\})\rightarrow 0 \mbox{ as }n\rightarrow +\infty,
\end{equation}
where the last inequality is due to the facts that $u_-(\cdot)$ is concave
and $u_-(0)=0$.

Now, define $\bar X_n:=nX-X_n$. Then $E[\rho\bar X_n]=nE[X\rho]-c_nE[\rho\id_{\rho>n^2}]=x_0$.
Moreover, since $\bar X_n^+\geq nX$ and $\bar X_n^-\leq c_n^+\id_{\rho>n^2}$,
it follows from (\ref{vxn0}) and (\ref{vxn}) that
$V(\bar X_n)\geq V_+(nX)-V_-(c_n^+\id_{\rho>n^2})\rightarrow +\infty$
as $n\rightarrow +\infty$.
\eof

\begin{remark}\label{rem4}
{\rm Quite intriguingly,
Theorem \ref{ip-2} shows that a probability distortion on {\it losses}
is {\it necessary} for the well-posedness if the utility on gains
can go arbitrarily large (the latter being the case
for most commonly used utility functions).
The intuition behind this result and its proof
can be explained as follows: one
borrows enormous amount of money to purchase a claim with a huge payoff
($nX$ in the proof), and then bet the market be ``good'' leading to
the realization of that payoff. If, for the lack of luck, the market
turns out to be ``bad'', then the agent ends up with a loss ($X_n$);
however due to the non-distortion on the loss side its
damage on value is bounded [in fact equation (10) shows that the damage can be controlled to be arbitrarily small]. Notice that the
above argument is no longer valid
if the wealth is {\it constrained} to be bounded from below\footnote{This is
why in Berkelaar {\it et al.} (2004) the model is well-posed even though
no probability distortion is considered, as the wealth process
there is constrained to be non-negative.}.}
\end{remark}

Now we set out to identify and solve well-posed problems.

\section{Main Results}

The original problem (\ref{btps}) is solved in two steps involving
{three} sub-problems, which are described in what follows.

\medskip

\noindent{\it Step 1}. In this step we consider two problems respectively:
\begin{itemize}
\item {\it Positive Part Problem}: A problem with parameters
$(A,x_+)$:
\begin{equation}\label{+part}
\begin{array}{ll}
\mbox{\rm Maximize} & V_+(X)=\int_0^{+\infty}T_+(P\{u_+(X)>y\})dy\\
\mbox{\rm subject to} & 
E[\rho X]=x_+,\;\;
               X\ge 0\;\as,\;\;
               X=0\;\as \mbox{ on } A^C,
\end{array}
\end{equation}
where $x_+\ge x_0^+\;(\geq 0)$ and $A\in \cF_T$ 
are given.
Thanks to Assumption \ref{ill}, $V_+(X)$ is a finite number for any feasible
$X$.
We define the optimal value of Problem (\ref{+part}), denoted
$v_+(A,x_+)$, in the following way. If
$P(A)>0$, in which case
the feasible region of (\ref{+part}) is non-empty [$X=(x_+\id_A)/(\rho P(A))$
is a feasible solution], then $v_+(A,x_+)$ is
defined to be the supremum of (\ref{+part}).
If $P(A)=0$ and $x_+=0$, then (\ref{+part}) has only one feasible solution $X=0\; \as$ and $v_+(A,x_+):=0$.
If $P(A)=0$ and $x_+>0$, then (\ref{+part}) has no
feasible solution, where we define $v_+(A,x):=-\infty$.

\item {\it Negative Part Problem}:
A problem with parameters
$(A,x_+)$:
\begin{equation}\label{-part}
\begin{array}{ll}
\mbox{\rm Minimize} & V_-(X)=\int_0^{+\infty}T_-(P\{u_-(X)>y\})dy\\
\mbox{\rm subject to} & \left\{\begin{array}{l}
E[\rho X]=x_+-x_0,\;
               X\ge 0\;\as,\;
               X=0\;\as \mbox{ on } A,\\
               X \mbox{ is upper bounded } \as,
      \end{array}\right.
\end{array}
\end{equation}
where $x_+\ge x_0^+$ and $A\in \cF_T$ 
are given.
Similarly to the positive part problem we define
the optimal value $v_-(A,x_+)$ of Problem (\ref{-part}) as follows.
When $P(A)<1$ in which case
the feasible region of (\ref{-part}) is non-empty, $v_-(A,x_+)$
is the infimum of (\ref{-part}).
If $P(A)=1$ and $x_+=x_0$ where the only feasible solution is $X=0\; \as$, then
$v_-(A,x_+):=0$.
If $P(A)=1$ and $x_+\neq x_0$, then there is no feasible solution, in which case
we define $v_-(A, x_+):=+\infty$.
\end{itemize}

\noindent{\it Step 2}. In this step we solve
\begin{equation}\label{step2}
\begin{array}{ll}
\mbox{\rm Maximize} &v_+(A,x_+)-v_-(A,x_+)\\
\mbox{\rm subject to}&\left\{\begin{array}{l}A\in \cF_T,\; x_+\ge x_0^+,\\
 \; x_+=0 \mbox{ when } P(A)=0,\;\; x_+=x_0 \mbox{ when } P(A)=1.
\end{array}\right.
\end{array}
\end{equation}


Let $F(\cdot)$ be the distribution function of $\rho$. Our main results
are stated in terms of the following mathematical program, which is intimately
related to (but not the same as) Problem (\ref{step2}):
\begin{equation}\label{simstep2}
\begin{array}{ll}
\mbox{\rm Maximize}& v_+(c, x_+)-u_-(\frac{x_+-x_0}{E[\rho\id_{\rho>c}]})T_-(1-F(c))\\
\mbox{\rm subject to} & \left\{\begin{array}{l}
                         \underline{\rho}\le c\le \bar\rho,\;\;x_+\ge x_0^+,\\
\;x_+=0 \mbox{ when } c=\underline{\rho},\;\;x_+=x_0 \mbox{ when } c=\bar \rho,
                         \end{array}\right.
\end{array}
\end{equation}
where $v_+(c, x_+):=v_+(\{\omega: \rho\le c\},x_+)$ and
we use the following convention:
\begin{equation}\label{convention}
u_-\left(\frac{x_+-x_0}{E[\rho\id_{\rho>c}]}\right)T_-(1-F(c)):=0 \;\; \mbox{ when }c=\bar\rho\mbox{ and }x_+=x_0.
\end{equation}

Here go the main results of this paper.

\begin{theo}\label{2stepeq3} Assume that $u_-(\cdot)$ is strictly concave at $0$.
We have the following conclusions:
\begin{itemize}
  \item [{\rm (i)}]  If
$X^*$ is optimal for Problem (\ref{btps}), then
$c^*:=F^{-1}(P\{X^*\ge 0\})$, $x_+^*:=E[\rho (X^*)^+]$,
where $F$ is the distribution function of $\rho$,
are optimal for Problem (\ref{simstep2}).
Moreover,  $\{\omega: X^*\ge 0\}$ and
            $\{\omega: \rho\le c^*\}$
are identical up to a zero probability set,
and $(X^*)^-=\frac{x_+^*-x_0}{E[\rho\id_{\rho>c^*}]}\id_{\rho>c^*}\;\; \as.$
\item [{\rm (ii)}] If $(c^*,x_+^*)$ is optimal for
Problem (\ref{simstep2}) and $X^*_+$
  is optimal for Problem (\ref{+part})
with parameters $(\{\rho\le c^*\},x_+^*)$,
then $X^*:=(X^*)^+\id_{\rho\leq c^*}-\frac{x_+^*-x_0}{E[\rho\id_{\rho>c^*}]}\id_{\rho>c^*}$
  is optimal for Problem (\ref{btps}).
  \end{itemize}
\end{theo}

In the light of Theorem \ref{2stepeq3}, we have the following algorithm to
solve Problem (\ref{btps}).

\begin{itemize}
\item[{\bf Step 1}] Solve Problem (\ref{+part}) with
$(\{\omega: \rho\le c\},x_+)$,
where $\underline{\rho}\le c\le \bar\rho$ and $x_+\ge x_0^+$ are given,
to obtain $v_+(c, x_+)$ and the optimal solution $X^*_+(c, x_+)$. 
\item[{\bf Step 2.}] Solve Problem (\ref{simstep2}) to get $(c^*,x_+^*)$.
\item[{\bf Step 3.}]
\begin{itemize}
\item[(i)] If $(c^*,x_+^*)=(\bar\rho, x_0)$,
then $X^*_+(\bar\rho ,x_0)$ solves Problem (\ref{btps}).
\item[(ii)] Else 
$X^*_+(c^*,x_+^*)\id_{\rho\le c^*}-\frac{x_+^*-x_0}{E[\rho\id_{\rho>c^*}]}\id_{\rho>c^*}$ solves
Problem (\ref{btps}).
\end{itemize}
\end{itemize}

We now impose the following assumption:
\begin{assump}\label{Fdec}
{\rm $F^{-1}(z)/T_+'(z)$ is non-decreasing in $z\in (0,1]$,
$\liminf_{x\rightarrow +\infty}\left(\frac{-xu_+^\dpm(x)}{u'_+(x)}\right)>0$,
and
 $E\left[u_+\left((u'_+)^{-1}(\frac{\rho}{T_+'(F(\rho))})\right)T_+'(F(\rho))\right]<+\infty$.}
\end{assump}
Then $v_+(c, x_+)$ and
the corresponding optimal solution $X_+^*$ to (\ref{+part}) can be
expressed more explicitly:
$$
\begin{array}{l}
v_+(c, x_+)=E\left[u_+\left((u_+')^{-1}\left(\frac{\lambda(c, x_+)\rho}{T_+'(F(\rho))}\right)\right)T_+'(F(\rho))\id_{\rho\le c}\right],\\
X^*_+=(u_+')^{-1}\left(\frac{\lambda(c, x_+)\rho}{T_+'(F(\rho))}\right)\id_{\rho\le c},
\end{array}
$$
where $\lambda(c, x_+)$ satisfies $E[(u_+')^{-1}(\frac{\lambda(c, x_+)\rho}{T_+'(F(\rho))})\rho\id_{\rho\le c}]=x_+$.
In this case Theorem \ref{2stepeq3} can be re-stated with the preceding
explicit expressions properly substituted.

{\rm Under Assumption \ref{Fdec}, the optimal terminal wealth
to our behavioral model (\ref{btps0}) is given
explicitly as the following
\begin{equation}\label{grandsolution}
X^*=(u_+')^{-1}\left(\frac{\lambda(c^*, x_+^*)\rho}{T_+'(F(\rho))}\right)\id_{\rho\leq c^*}-\frac{x_+^*-x_0}{E[\rho\id_{\rho>c^*}]}\id_{\rho>c^*}.
\end{equation}
This solution possesses some appealing features.
On one hand, the terminal wealth having a gain or a loss is completely
determined by
the terminal state density price being lower or higher than a single threshold,
$c^*$, which in turn can be obtained by
solving (\ref{simstep2}).
On the other hand, (\ref{grandsolution}) is
the payoff of a combination of two binary options, which can be easily
priced; see Appendix \ref{replicate}.}

The remainder of this paper is devoted to proving all the above claims. But before that, let us discuss on the
economical interpretation of the optimal wealth profile (\ref{grandsolution}).
Indeed, (\ref{grandsolution}) suggests that
an optimal strategy should deliver
a wealth in excess of the reference wealth in good states of the world
($\rho\leq c^*$), and a shortfall in bad states ($\rho>c^*$)\footnote{It can be easily shown,
in the case of a one-stock market, that $\rho\leq c^*$ is equivalent to
the stock price exceeding a certain level.}.
To realize this goal, the agent should initially
buy a contingent claim with the payoff
$(u_+')^{-1}\left(\frac{\lambda(c^*, x_+^*)\rho}{T_+'(F(\rho))}\right)\id_{\rho\leq c^*}$ at cost $x^*_+$. Since $x^*_+\geq x_0$, he
needs to issue (i.e., sell)
a claim with a payoff $\frac{x_+^*-x_0}{E[\rho\id_{\rho>c^*}]}\id_{\rho>c^*}$ to finance the shortfall, $x^*_+-x_0$. In other words, the agent will not only invest in stocks, but also will generally take a {\it leverage} to do so. He then gambles on
a good state of the market turning up at the terminal time while accepting
a {\it fixed} loss in case of a bad state\footnote{Such a gambling
policy was derived in Berkelaar {\it et al.} (2004), Proposition 3,
for a special model where the value function is a two-piece power function
and there is no probability distortion. Here, we show that even for the most
general model an optimal behavioral policy still possesses such an elegantly
simple structure.}.

\section{Splitting}

The key idea developed in this paper, i.e., splitting (\ref{btps}) into
three sub-problems and then appropriately merging them,
is based on the following observation:
If $X$ is a feasible solution of (\ref{btps}), then one can split
$X^+$ and $X^-$. The former defines naturally an event
$A:=\{X\ge0\}$ and an initial price $x_+:=E[\rho X^+]$, and the latter
corresponds to $A^C$ and $x_+-x_0$,
where $A^C$ denotes the complement of the set $A$.
An optimal solution to (\ref{btps})
should, therefore, induce the ``best'' such $A$ and $x_+$ in certain sense.
We now prove that this idea indeed works in the sense that (\ref{btps}) is
equivalent to the three auxiliary problems combined.

We start with the well-posedness.
\begin{prop}\label{ill=ill}
{Problem (\ref{btps}) is ill-posed if and only if Problem (\ref{step2}) is ill-posed.}
\end{prop}
{\sc Proof:} We first show the ``if'' part.
Suppose (\ref{step2}) is ill-posed. If $v_+(\Omega, x_0)=+\infty$, then Problem (\ref{btps}) is obviously
ill-posed. If $v_+(\Omega, x_0)<+\infty$,  then for any $M>v_+(\Omega, x_0)$, there exists a feasible
pair $(A,x_+)$ for (\ref{step2}) such that $v_+(A,x_+)-v_-(A,x_+)\ge M$.
Clearly $0<P(A)<1$.
(If $P(A)=0$, then $v_+(A,x_+)-v_-(A,x_+)\le 0<M$. If
$P(A)=1$, then $v_+(A,x_+)-v_-(A,x_+)\le v_+(\Omega, x_0)<M$.)
Consequently,
both (\ref{+part}) and (\ref{-part}) with parameters $(A, x_+)$, $x_+\ge x_0^+$, have non-empty feasible regions.
So there exist $X_1$ and $X_2$ feasible for (\ref{+part})
and (\ref{-part}) respectively such that $V_+(X_1)\ge v_+(A,x_+)-1, V_-(X_2)\le v_-(A,x_+)+1$.
Define $X=X_1-X_2$. Then $X$ is feasible for (\ref{btps}), and $V(X)\ge v_+(A,x_+)-v_-(A,x_+)-2\ge M-2$,
implying that (\ref{step2}) is ill-posed.

For the ``only if'' part, if (\ref{btps}) is ill-posed, then for any $M>0$, there exists a feasible
solution $X$ for (\ref{btps}) such that $V(X)\ge M$. Define $A:=\{\omega: X\ge 0\},x_+:=E[\rho X^+]$. Then
$(A, x_+)$ is feasible for Problem (\ref{step2}), and
$v_+(A,x_+)-v_-(A, x_+)\ge V(X^+)-V(X_-)=V(X)\ge M$, which shows that Problem (\ref{step2}) is ill-posed.
\eof

\begin{prop}\label{2stepeq}
Given $X^*$, define $A^*:=\{\omega: X^*\ge 0\}$ and $x_+^*:=E[\rho(X^*)^+]$.
Then $X^*$  is optimal for Problem (\ref{btps}) if and only if
$(A^*,x_+^*)$ are optimal for Problem (\ref{step2}) and  $(X^*)^+$ and $(X^*)^-$ are respectively optimal
for Problems (\ref{+part}) and (\ref{-part})  with parameters $(A^*, x_+^*)$.
\end{prop}

{\sc Proof:}
For the ``if'' part, we first have $V(X^*)=v_+(A^*, x_+^*)-v_-(A^*,x_+^*)$.
For any feasible solution $X$ of
 (\ref{btps}),
define $A:=\{\omega: X\ge 0\}$ and $x_+:=E[\rho X^+]$. Then we have $V_+(X^+)\le v_+(A,x_+), V_-(X^-)\ge v_-(A, x_+)$.
Therefore $V(X)=V_+(X^+)-V_-(X^-)\le v_+(A,x_+)-v_-(A,x_+)\le v_+(A^*, x_+^*)-v_-(A^*,x_+^*)=V(X^*)$, which means
$X^*$ is optimal for (\ref{btps}).

For the ``only if'' part, let $X^*$ be optimal for (\ref{btps}).
   Obviously, $V_+((X^*)^+)\le v_+(A^*, x_+^*)$ and $V_-((X^*)^-)\ge v_-(A^*, x_+^*)$.
   If the former holds strictly, then there exists $X_1$ feasible for (\ref{+part})
   with parameters $(A^*,x_+^*)$ such that $V_+(X_1)> V_+((X^*)^+)$.
As a result $\bar X:=X_1\id_{A^*}+X^*\id_{(A^*)^C}$
   is feasible for (\ref{btps}) and $V(\bar X)>V(X^*)$, which contradicts the optimality of $X^*$.
   So $(X^*)^+$ is optimal for (\ref{+part}). Similarly we can prove that
$(X^*)^-$ is optimal for (\ref{-part}). Thus
   $v_+(A^*,x_+^*)=V_+((X^*)^+),v_-(A^*,x_+^*)=V_-((X^*)^-)$.

Next we show that $v_+(A, x_+)-v_-(A,x_+)\le v_+(A^*, x_+^*)-v_-(A^*,x_+^*)
\equiv V(X^*)$
for any feasible pair $(A,x_+)$ of Problem (\ref{step2}). This can be proved in three cases:
\begin{itemize}
  \item [(i)] If $P(A)=0$ (hence $x_+=0$ and $x_0\le 0$), then
  \begin{eqnarray*}
v_+(A, x_+)-v_-(A,x_+)&=&-v_-(A,0)\\
&=& -v_-(A,x_0^+)\\
&=&\sup_{E[\rho X]=x_0^-,\;X\ge 0,\;X \mbox{ is upper bounded }}[-V_-(X)]\\
&=&\sup_{E[\rho X]=-x_0^-,\;X\le 0,\;X \mbox{ is lower bounded }}V(X)\\
&\le&\sup_{E[\rho X]=-x_0^-,\;X \mbox{ is lower bounded }}V(X)\\
&= &V(X^*),
 \end{eqnarray*}
where the last equality is owing to the fact that $-x_0^-= x_0$.

  \item [(ii)] If $P(A)=1$ (hence $x_+=x_0$), then we need only to check $v_+(A, x_0)\le V(X^*)$, which is
  easy since $v_+(A,x_0)=\sup_{E[\rho X]=x_0,\;X\ge 0}V(X)$.

  \item [(iii)] If $0<P(A)<1$, then for any $x_+\ge x_0^+$, both
(\ref{+part}) and (\ref{-part}) with parameters
$(A,x_+)$ have non-empty feasible regions.
Hence for any $\epsilon>0$
there exist $X_1$ and $X_2$, feasible for (\ref{+part}) and (\ref{-part})
  respectively, such that $V_+(X_1)>(v_+(A,x_+)-\epsilon), V_-(X_2)<v_-(A,x_+)+
\epsilon$.
Letting $X:=X_1-X_2$, which is feasible for (\ref{btps}), we have
$v_+(A, x_+)-v_-(A,x_+)<V_+(X_1)-V_-(X_2)+2\epsilon
=V(X)+2\epsilon\le V(X^*)+2\epsilon$.
\end{itemize}

This concludes the proof.
\eof

The essential message of
Propositions \ref{ill=ill} and \ref{2stepeq} is that
our problem (\ref{btps}) is completely equivalent to the set of problems
(\ref{+part}) --  (\ref{step2}) and, moreover,
the solution to the former  can be obtained via those to the latter.

Problem (\ref{step2}) is an optimization problem with
the decision variables being a real number, $x_+$, and a random event, $A$,
the latter being very hard to handle. We now show that
one  needs only to consider $A=\{\rho\le c\}$, where $c$ is a real number in
certain range,
when optimizing (\ref{step2}).

Recall that two random variables $\xi$ and $\eta$ are called {\it comonotonic}
({\it anti-comonotonic} respectively) if
$[\xi(\omega)-\xi(\omega')][\eta(\omega)-\eta(\omega')]\geq\; \mbox{($\leq$
respectively) }\;0$.

\begin{theo}\label{onA}
 For any feasible pair $(A, x_+)$ of Problem (\ref{step2}), 
 there exists $c\in [\underline{\rho} , \bar\rho ]$ such that
$\bar A:=\{\omega: \rho\le c\}$ satisfies
\begin{equation}\label{vdiff}
v_+(\bar A, x_+)-v_-(\bar A,x_+)\ge v_+(A, x_+)-v_-(A,x_+).
\end{equation}
Moreover, if Problem (\ref{+part}) admits an optimal solution
with parameters $(A, x_+)$,
then the inequality in (\ref{vdiff})
is strict unless $P(A\cap \bar A^C)+P(A^C\cap \bar A)=0$.
\end{theo}

{\sc Proof:}
The case when $x_+=x_0^+$ is trivial.
In fact, if $x_0\le 0$, then $x_+=0$ and $v_+(A, x_+)=0$ $\forall A$;
hence $c=\underline{\rho}$ or $\bar A=\O$.
If $x_0>0$, then obviously
$v_-(A,x_+)=0$ $\forall A$; hence (\ref{vdiff}) holds with
$\bar A=\Omega$ or $c=\bar \rho$.
On the other hand,
the case when $P(A)=0$  or $P(A)=1$ is also trivial, where $c:=\underline{\rho} $ or $c:=\bar \rho$
trivially meets (\ref{vdiff}).

So we assume now that $x_+>x_0^+$ and $0<P(A)<1$.
Denote $\alpha:=P(A)$, $B:=A^C$.
Let $\bar A=\{\omega: \rho\le c\}$, where
$c\in [\underline{\rho}, \bar\rho )$ satisfies
$P\{\rho\le c\}=\alpha$.
Further, set
$$
\begin{array}{ll}
A_1=A\cap\{\omega: \rho\le c\},& A_2=A\cap\{\omega: \rho>c\},\\
B_1=B\cap\{\omega: \rho\le c\},& B_2=B\cap\{\omega: \rho>c\}.
\end{array}
$$
Since $P(A_1\cup B_1)=P(A_1\cup A_2)\equiv \alpha$, we conclude
$P(A_2)=P(B_1)$.

If $P(A_2)=P(B_1)=0$, then 
{trivially} $v_+(\bar A, x_+)-v_-(\bar A,x_+)= v_+(A, x_+)-v_-(A,x_+)$.
So we suppose $P(A_2)=P(B_1)>0$.
For any feasible solutions $X_1$ and $X_2$ for (\ref{+part}) and (\ref{-part}),
respectively, with parameters $(A,x_+)$, we are to prove that
\begin{equation}\label{fullimprov}
V_+(X_1)-V_-(X_2)\le v_+(\bar A, x_+)-v_-(\bar A, x_+).
\end{equation}
To this end, define $f_1(t):=P\{X_1\le t|A_2\}$, $g_1(t):=P\{\rho\le t|B_1\}$,
$t\geq0$,
$Z_1:=g_1(\rho)$ and $Y_1:=f_1^{-1}(Z_1)$.
Because $\rho$ admits no atom with respect to $P$,
it admits no atom with respect to $P(\cdot|B_1)$.
Hence
the distribution of $Z_1$ conditional on $B_1$ is $U(0,1)$, which
implies
$P\{Y_1\le t|B_1\}=P\{Z_1\leq f_1(t)|B_1\}=f_1(t)$. Consequently,
\begin{eqnarray*}
E[\rho X_1\id_{A_2}]\ge cE[X_1\id_{A_2}]&=&cP(A_2)E[X_1|A_2]\\
&=&cP(B_1)\int_0^{+\infty}[1-f_1(t)]dt\\
&=&cP(B_1)\int_0^{+\infty}P\{Y_1>t|B_1\}dt\\
&=&cE[Y_1\id_{B_1}]\\
&\ge&E[\rho Y_1\id_{B_1}],
\end{eqnarray*}
and the inequality is strict if and only if $P\{X_1>0\}>0$ or
$f_1(t)\not\equiv 1$.

Define
\[k_1:=\left\{
\begin{array}{ll}
1, &\mbox{ if } Y_1=0,\;\as{\mbox{ on } B_1},\\
\frac{E[\rho X_1\id_{A_2}]}{E[\rho Y_1\id_{B_1}]},& \mbox{ otherwise. }
\end{array}\right.
\]
Then $k_1\ge 1$, and $k_1>1$ if and only if $f_1(t)\not\equiv 1$. Set
$\bar X_1:=X_1\id_{A_1}+k_1Y_1\id_{B_1}$. Then
$$E[\rho X_1]=E[\rho X_1\id_{A_1}]+E[\rho X_1\id_{A_2}]
=E[\rho X_1\id_{A_1}]+E[k_1\rho Y_1\id_{B_1}]=E[\rho\bar X_1],$$
which means that $\bar X_1$ is feasible for (\ref{+part}) with
parameters $(\bar A,x_+)$ (recall that by definition $\bar X_1=0$ on
$\bar A^C$).

On the other hand, for any $t{>} 0$,
\begin{eqnarray*}
  P\{\bar X_1>t\}&=&P\{\bar X_1>t|A_1\}P(A_1)+P\{\bar X_1>t|B_1\}P(B_1)\\
  &=&P\{X_1>t|A_1\}P(A_1)+P\{k_1 Y_1>t|B_1\}P(B_1)\\
  &\ge&P\{X_1>t|A_1\}P(A_1)+P\{Y_1>t|B_1\}P(B_1)\\
  &=&P\{X_1>t|A_1\}P(A_1)+P\{X_1>t|A_2\}P(A_2)\\
  &=&P\{X_1>t\},
\end{eqnarray*}
and the equality holds {for $\aee\; t\ge 0$}  
if and only if $f_1(t)\equiv 1$.

It follows then by the definition of $V_+(\cdot)$ that
\begin{equation}\label{+improv}
V_+(\bar X_1)\ge V_+(X_1),
\end{equation}
with the inequality being strict when $f_1(\cdot)\not\equiv 1$.

In a similar fashion
we can construct $\bar X_2$
feasible for (\ref{-part}) with parameters $(\bar A,x_+)$ satisfying
\begin{equation}\label{-improv}
V_-(\bar X_2)\le V_-(X_2).
\end{equation}
Combining (\ref{+improv}) and (\ref{-improv}) we get (\ref{fullimprov}).

Now, if $X_1$ is an optimal solution of (\ref{+part})
with parameters $(A, x_+)$,
then $P(X_1=0|A_2)<1$. Indeed, if $P(X_1=0|A_2)=1$,
then by its optimality $X_1$ is anti-comonotonic
with $\rho$ on $A$ (see Proposition \ref{maxv+-a}), which implies $P(X_1=0|A_1)=1$.
Therefore $P(X_1=0|A)=1$, and $x_+=E[\rho X_1\id_A]=0$,
contradicting the fact that $x_+>x_0^+\geq 0$.

Thus, $f_1(\cdot)\not\equiv 1$. As proved earlier, (\ref{+improv}), and
hence (\ref{fullimprov}), holds
strictly.
\eof

To simplify the notation, we now use $v_+(c, x_+)$ and $v_-(c, x_+)$
to denote $v_+(\{\omega: \rho\le c\},x_+)$ and
$v_-(\{\omega: \rho\le c\},x_+)$ respectively.

In view of Theorem \ref{onA},
one may replace  Problem (\ref{step2})
by the following problem:
 \begin{equation}\label{step2c}
\begin{array}{ll}
\mbox{\rm Maximize} &v_+(c,x_+)-v_-(c,x_+)\\
\mbox{\rm subject to }&\left\{\begin{array}{l} \underline{\rho}\leq c\leq \bar \rho,\;\; x_+\ge x_0^+,\\
\; x_+=0 \mbox{ when } c=\underline{\rho},\;\;x_+=x_0 \mbox{ when } c=\bar\rho.
               \end{array}\right.
\end{array}
\end{equation}
This is clearly a {\it much} simpler problem, being a constrained
optimization problem (a mathematical programming problem) in $\R^2$.

{\rm Theorem \ref{onA} is one of the most important results in this paper.
It discloses the form of a general solution to the behavioral model:
the optimal wealth
is the payoff of a combination of two binary options characterized by a single number
$c^*$, as stipulated in the next theorem.}

\begin{theo}\label{2stepeq2}
  Given $X^*$, and define $c^*:=F^{-1}(P\{X^*\ge 0\})$, $x_+^*:=E[\rho (X^*)^+]$, where $F(\cdot)$ is the distribution function of $\rho$.
  Then $X^*$ is optimal for Problem (\ref{btps}) if and only if $(c^*, x_+^*)$ is optimal for Problem (\ref{step2c})
and
  $(X^*)^+\id_{\rho\le c^*}$ and $(X^*)^-\id_{\rho>c^*}$ are respectively
optimal for Problems
(\ref{+part}) and (\ref{-part}) with parameters $(\{\omega: \rho\le c^*\},x_+^*)$. Moreover, in this case
  $\{\omega: X^*\ge 0\}$ and $\{\omega: \rho\le c^*\}$ are identical up to a zero probability set.
\end{theo}

{\sc Proof:} Straightforward from Proposition \ref{2stepeq} and Theorem \ref{onA}.
\eof

In the following two sections, we will solve
the positive and negative part problems respectively
to obtain
$v_+(c, x_+)$ and $v_-(c, x_+)$.
It turns out that the two problems require very different techniques to
tackle.

\section{Positive Part Problem}

In this section we solve the positive part problem (\ref{+part}), including
finding its optimal solution and the expression of $v_+(c, x_+)$,
for any $A=\{\omega: \rho\le c\}$, $\underline{\rho}\leq c\le \bar \rho$, and $x_+\geq x_0^+$.
In fact, it is a special case of a more general Choquet maximization
problem, which is of independent interest and is solved in Appendix \ref{pppsubsection}.

\subsection{Solving (\ref{+part})}

We apply the general results obtained in Appendix \ref{pppsubsection} to Problem (\ref{+part}) with $A=\{\omega: \rho\le c\}$
and $x_+\ge x_0^+\; (\ge 0)$.
Let $F(\cdot)$ be the distribution function of $\rho$.

Let $A=\{\omega: \rho\le c\}$ be given.
Problem (\ref{+part}) is trivial when $P(A)=0$; hence we assume
$P(A)>0$ or $c>\underline{\rho}$. Define
\[ T_A(x):=T_+(xP(A))/T_+(P(A)),\;\;x\in[0,1],
\]
which is a strictly increasing, differentiable function from $[0,1]$ to $[0,1]$,
 with $T_A(0)=0, T_A(1)=1$. For any feasible
solution $X$ of (\ref{+part}) and any $y\geq0$,
$$T_+\left(P\{u_+(X)>y\}\right)=T_+\left(P\{u_+(X)>y|A\}P(A)\right)=
T_+(P(A))T_A\left(P\{u_+(X)>y|A\}\right).$$
Now considering Problem (\ref{+part})
in the conditional probability space $\left(\Omega\cap A, \cF\cap A, P_A:=P(\cdot|A)\right)$,
we can rewrite it as
\begin{equation}\label{+part2}
\begin{array}{ll}
\mbox{\rm Maximize}& V_+(Y)=T_+(P(A))\int_0^{+\infty} T_A(P_A\{u_+(Y)>y\})dy\\
\mbox{\rm subject to} & E_A[\rho Y]=x_+/P(A), \;\; Y\ge 0.
\end{array}
\end{equation}
This specializes the general Choquet maximization problem (\ref{plusutility})
solved in Appendix \ref{pppsubsection}.
It is evident that $Y^*$ is optimal for (\ref{+part2}) if and only if $X^*=Y^*\id_A$ is
optimal for (\ref{+part}).

To solve Problem (\ref{+part}) for all $A=\{\omega: \rho\le c\}$, we
need Assumption \ref{Fdec}.

\begin{theo}\label{os+p2}
Let Assumption \ref{Fdec} hold. Given $A:=\{\omega: \rho\le c\}$ with
$\underline{\rho}\leq c\le \bar\rho$, and $x_+\ge x_0^+$.
\begin{itemize}
\item[{\rm (i)}] If $x_+=0$, then the optimal solution of (\ref{+part}) is $X^*=0$ and $v_+(c, x_+)=0$.
\item[{\rm (ii)}] If $x_+>0$ and $c=\underline{\rho} $, then
there is no feasible solution to (\ref{+part}) and $v_+(c, x_+)=-\infty$.
\item[{\rm (iii)}] If $x_+>0$ and
$\underline{\rho} <c\le \bar\rho $,
then the optimal solution to (\ref{+part}) is
$X^*(\lambda)=(u'_+)^{-1}\left(\frac{\lambda\rho}{T_+'(F(\rho))}\right)\id_{\rho\le c}$
with the optimal value
$v_+(c,x_+)=E\left[u_+\left((u_+')^{-1}(\frac{\lambda\rho}{T_+'(F(\rho))})\right)T_+'(F(\rho))\id_{\rho\le c}\right]$, where
$\lambda>0$ is the unique real number satisfying $E[\rho X^*(\lambda)]=x_+$.
\end{itemize}
\end{theo}

{\sc Proof:}
Cases (i) and (ii) are trivial. We prove (iii).
Assume $\underline{\rho} <c\le \bar\rho $ with $P(A)\equiv P\{\rho\le c\}>0$.
Define $F_A(x):=P_A\{\rho\le x\}=\frac{P\{\rho\le x\wedge c\}}{P\{\rho\le c\}}=\frac{F(x\wedge c)}{P(A)}$, $x\geq0$.
Then $F_A^{-1}(x)=F^{-1}(xP(A))$.
Noting $T_A'(x)=\frac{P(A)}{T_+(P(A))}T_+'(xP(A))$, we have
$\frac{F_A^{-1}(z)}{T_A'(z)}=\frac{F^{-1}(zP(A))}{T_+'(zP(A))}\frac{T_+(P(A))}{P(A)}$, which is
non-decreasing in $z$ under Assumption \ref{Fdec}.
Noting that $\rho\le c$ on $A$, we have
$\frac{\rho}{T_A'(F_A(\rho))}=\frac{\rho}{T_+'(F(\rho))}\frac{T_+(P(A))}{P(A)}$.
Hence, in view of Assumption \ref{Fdec} and Proposition \ref{lagequiv}
we can apply
Theorem \ref{maxv} to conclude that the optimal solution for
(\ref{+part2}) is
$Y^*=(u_+')^{-1}\left(\frac{\bar \lambda \rho}{T_A'(F_A(\rho))}\right)$ for some $\bar \lambda> 0$.
Denoting $\lambda:=\frac{T_+(P(A))}{P(A)}\bar \lambda\geq 0$,
we obtain the optimality of $X^*:=Y^*\id_{\rho\le c}$
in view of the relation between Problems
(\ref{+part2}) and (\ref{+part}).

Finally, the optimal value of Problem (\ref{+part}) can be calculated as follows:
\begin{eqnarray*}
v_+(c, x_+)&=&T_+(P(A))E_A[u_+(Y^*)T_A'(F_A(\rho))]\\
&=&P(A)E_A[u_+(Y^*)T_+'(F(\rho))]\\
&=& E[u_+(Y^*)T_+'(F(\rho))\id_{\rho\le c}].
\end{eqnarray*}
The proof is complete.
\eof

{\rm Theorem \ref{os+p2} remains true when the condition $\liminf_{x\rightarrow +\infty}\left(\frac{-xu_+^\dpm(x)}{u'_+(x)}\right)>0$ in Assumption \ref{Fdec}
is replaced by a (mathematically) weaker one
$$\limsup_{x\rightarrow +\infty}\frac{u'_+(kx)}{u'_+(x)}<1 \mbox{  for some }
k>1,$$
which,
in particular, does not require the twice differentiability of $u_+(\cdot)$;
see Jin, Xu and Zhou (2007, Lemma 3 and Proposition 2).
We choose to use the current condition due to its clear economic meaning related to the relative risk aversion index.}

Before we end this subsection, we state the following result which is
useful in the sequel.

\begin{prop}\label{strictincofV+}
  If $x_+>0$, 
  then
Problem (\ref{+part}) admits an optimal solution with
parameters $(\{\rho\leq c\},x_+)$ only if
  $v_+(\bar c, x_+)>v_+(c, x_+)$ for any $\bar c> c$
satisfying $P\{c<\rho\leq\bar c\}>0$.
\end{prop}

{\sc Proof:} Assume $c>\underline{\rho}$, the case $c=\underline{\rho}$ being trivial.
Let $X$ be optimal for (\ref{+part}) with $(A(c), x_+)$,
where $A(c):=\{\omega: \rho\le c\}$.
Then $Y_c:=X|_{A(c)}$, where $X|_{A(c)}$ is $X$ restricted on $A(c)$,
is optimal for
(\ref{+part2}) with $A=A(c)$.

For any $\bar c>c$,
obviously $v_+(\bar c, x_+)\ge v_+(c, x_+)$.
If $v_+(\bar c,x_+)=v_+(c, x_+)$, then,
with $A(\bar c):=\{\omega: \rho\le \bar c\}$, the random
variable
$$\bar Y(\omega):=\left\{\begin{array}{ll}Y_c(\omega),& \mbox{ if }\omega\in A(c),\\
 0,&\mbox{ if }\omega\in A(\bar c)\setminus A(c)\end{array}\right.$$
is feasible for (\ref{+part2}) with $A=A(\bar c)$ and, since
its objective value
is $v_+(c, x_+)=v_+(\bar c, x_+)$, is optimal.
By Theorem \ref{+optdis}, $P\{\bar Y=0|A(\bar c)\}=0$.
However, the definition of $\bar Y$ shows that
$P\{\bar Y=0|A(\bar c)\}>0$ if $P\{c<\rho\leq\bar c\}>0$.
This contradiction leads to $v_+(\bar c, x_+)>v_+(c, x_+)$. \eof

In other words, $v_+$ is {\it strictly} increasing in $c$.

\subsection{Discussion on the Monotonicity of $F^{-1}(z)/T_+'(z)$}
It is seen from the previous subsections that
in order to solve the positive part problem {\it explicitly},
a key assumption is the monotonicity of $F^{-1}(z)/T_+'(z)$.
What is the economic interpretation of this property?
Does it contradict the other assumptions usually imposed on $F(\cdot)$
and $T_+(\cdot)$? More importantly, is the set of the problem parameters
satisfying this assumption null in the first place?
In this subsection we depart from our optimization problems
for a while to address these questions\footnote{The reader may skip
this subsection without interrupting the flow of reading.}.

Throughout this subsection, we assume that $F(\cdot)$ (the distribution
function of $\rho$) is twice
differentiable and $F'(x)>0$ $\forall x>0$ (e.g., when $\rho$ is
a non-degenerate lognormal random variable).
{Furthermore, suppose that $T_+(\cdot)$ is twice differentiable on (0,1).}

Denote $x=F^{-1}(z)$ or $z=F(x)$. Then
the monotonicity (being non-decreasing) of $F^{-1}(z)/T_+'(z)$ is equivalent to
that $T_+'(F(x))/x$ is non-increasing in $x>0$.
Set $H(x):=T_+(F(x)), h(x):=H'(x)$,
and
$I(x):=T_+'(F(x))/x\equiv h(x)/(xF'(x))$, $x>0$.
Then $I(x)$ non-increases in $x>0$ if and only if
\[I'(x)=\frac{xH^\dpm(x)F'(x)-xH'(x)F^\dpm(x)-H'(x)F'(x)}{x^2 (F'(x))^2}\le 0\;\;\forall x>0,
\]
which is further equivalent to
\begin{equation}\label{equality1}
\frac{xH^\dpm(x)}{H'(x)}- \frac{xF^\dpm(x)}{F'(x)}\le 1\;\;\forall x>0
\end{equation}
or
\begin{equation}\label{equality2}
(\ln H'(x))'\le (\ln( xF'(x)))'\;\;\forall x>0.
\end{equation}

Note
that $\frac{xu^\dpm(x)}{u'(x)}$ can be regarded as the {\it relative
risk seeking index} of a given function $u(\cdot)$.
On the other hand, recall that by definition
$H(\cdot)$ is the {\it distorted} distribution function of
$\rho$. Hence the condition
(\ref{equality1}) can be economically interpreted as that
the distortion $T_+$ should not be ``too large'' in the sense that
it should not increase the relative risk seeking function
of the distribution by more than 1.

Next we are to explore more properties
of the function $j(\cdot)$ defined by
$j(x):=\frac{xH^\dpm(x)}{H'(x)}-\frac{xF^\dpm(x)}{F'(x)},\;\;x>0$.
To this end,
let $G(z):=F^{-1}(z)$. Then
$ G'(z)=\frac{1}{F'(G(z))}\;\;\forall z\in(0,1)$. 
Since $T_+(z)=H(G(z))$, we have $T_+'(z)=h(G(z))/F'(G(z))$; hence
\[ T_+^\dpm(z)=\frac{h'(G(z))F'(G(z))-h(G(z))F^\dpm(G(z))}{F'(G(z))^3}.\]
This leads to
\begin{equation}\label{t+dpm}
T_+^\dpm(F(x))=\frac{h'(x)F'(x)-h(x)F^\dpm(x)}{F'(x)^3}=
\frac{h(x)}{xF'(x)^2}j(x),\;x>0.
\end{equation}

As proposed by Tversky and Kahneman (1992),
the probability distortion $T_+(\cdot)$
is usually in {\it reversed} S-shape. Specifically, $T_+(x)$ changes from being
concave to being convex when $x$ goes from $0$ to $1$, or
$T_+^\dpm(x)$ changes from negative to positive.
It follows then from (\ref{t+dpm}) that
$j(\cdot)$ changes from negative to positive when $x$ goes from $0$ to $1$,
while as shown earlier (\ref{equality1}) requires that $j(\cdot)$ is bounded above by 1.

To summarize,
a reversed S-shaped distortion $T_+(\cdot)$ satisfying
the monotonicity condition in Assumption \ref{Fdec}
{\it if} there exists $c_0>0$ such that
\begin{equation}\label{gproperty}
j(x)\le 0\;\;\forall x\in (0, c_0], \;\;\mbox{ and }\;\; 0\le j(x)\le 1\;\;
\forall x\in (c_0, +\infty).
\end{equation}

The following is an example of distortion where the corresponding $j(\cdot)$
does satisfy (\ref{gproperty}).

\begin{exam}\label{T+}
{\rm Let $\rho$ be a non-degenerate lognormal random variable; i.e.,
$F(x)=N\left(\frac{\ln x-\mu}{\sigma}\right)$ for some $\mu\in\R$ and $\sigma>0$, where $N(\cdot)$ is the distribution function of
a standard normal random variable. Take $j(x)=:a\id_{0<x\le c_0}+b\id_{x>c_0}$, with $c_0>0$, $a<0$ and $0<b<1$ all
given. This is the ``simplest'' function satisfying (\ref{gproperty}).
We now track down the distortion $T_+(\cdot)$ that produces the
function $j(\cdot)$.

When $0<x\le c_0$,
$j(x)\equiv x\left[(\ln H'(x))'-(\ln F'(x))'\right]=a$.
Hence
\[\ln H'(x)-\ln F'(x)=\bar k+a\ln x\]
 for some constant $\bar k$, or
\begin{equation}\label{Ha}
  H'(x)=kF'(x)x^{a}=\frac{k}{\sqrt{2\pi}\sigma}x^{a-1}e^{-(\ln x-\mu)^2/(2\sigma^2)},\;\;0<x\le c_0,
\end{equation}
for some constant $k$.
Thus,
\begin{equation}\label{dis1}
\begin{array}{rcl}
H(x)&=&\frac{k}{\sqrt{2\pi}\sigma}\int_0^xt^{a-1}e^{-(\ln t-\mu)^2/(2\sigma^2)}dt\\
&=&\frac{k}{\sqrt{2\pi}\sigma}\int_{-\infty}^{\ln x}e^{as}e^{-(s-\mu)^2/(2\sigma^2)}ds\\
&=&\frac{k}{\sqrt{2\pi}\sigma}e^{a\mu+a^2\sigma^2/2}\int_{-\infty}^{\ln x}e^{-(s-(\mu+a\sigma^2))^2/(2\sigma^2)}ds\\
&=&ke^{a\mu+a^2\sigma^2/2}N\left(\frac{\ln x-(\mu+a\sigma^2)}{\sigma}\right),\;0<x\le c_0.
\end{array}
\end{equation}
Consequently,
\[ T_+(z)\equiv H(F^{-1}(z))=ke^{a\mu+a^2\sigma^2/2}N\left(
N^{-1}(z)-a\sigma\right),  \qquad  0<z \le F(c_0):=z_0.
\]

When $x>c_0$, similar to (\ref{Ha}) we have
\begin{equation}\label{Hb}
  H'(x)=\frac{\tilde k}{\sqrt{2\pi}\sigma}x^{b-1}e^{-(\ln x-\mu)^2/(2\sigma^2)},\;\;
x>c_0,
\end{equation}
with {$\tilde k=c_0^{a-b}k$ (to render $H'(x)$ continuous at $x=c_0$)}.
Therefore,
{\small \begin{equation}\label{dis2}
\begin{array}{rcl}
H(x)&=&H(c_0)+\frac{\tilde k}{\sqrt{2\pi}\sigma}\int_{c_0}^xt^{b-1}e^{-(\ln t-\mu)^2/(2\sigma^2)}dt\\
&=&H(c_0)+\frac{\tilde k}{\sqrt{2\pi}\sigma}\int_{\ln c_0}^{\ln x}e^{bs}e^{-(s-\mu)^2/(2\sigma^2)}ds\\
&=&H(c_0)+\frac{\tilde k}{\sqrt{2\pi}\sigma}e^{b\mu+b^2\sigma^2/2}\int_{\ln c_0}^{\ln x}e^{-(s-(\mu+b\sigma^2))^2/(2\sigma^2)}ds\\
&=&H(c_0)+\tilde ke^{b\mu+b^2\sigma^2/2}
   \left[N\left(\frac{\ln x-(\mu+b\sigma^2)}{\sigma}\right)-N\left(\frac{\ln c_0-(\mu+b\sigma^2)}{\sigma}\right)\right],\;x>c_0.
\end{array}
\end{equation}}
This leads to
\[ \begin{array}{rl}
&T_+(z)=H(F^{-1}(z))\\
=&ke^{a\mu+a^2\sigma^2/2}N\left(N^{-1}(z_0)-a\sigma\right)
  +\tilde ke^{b\mu+b^2\sigma^2/2}\left[N\left(N^{-1}(z)-b\sigma\right)
-N\left(N^{-1}(z_0)-a\sigma\right)\right],\\
&\;\;\;\;\;\;z_0<z\leq 1.
\end{array}
\]
In particular,
\[\begin{array}{rl}
 T_+(1)=&ke^{a\mu+a^2\sigma^2/2}N\left(N^{-1}(z_0)-a\sigma\right)
  +\tilde ke^{b\mu+b^2\sigma^2/2}\left[1-N\left(N^{-1}(z_0)-a\sigma\right)\right]\\
=&ke^{a\mu+a^2\sigma^2/2}N\left(\frac{\ln c_0-\mu-a\sigma^2}{\sigma}\right)
 +kc_0^{a-b}e^{b\mu+b^2\sigma^2/2}\left[1-N\left(\frac{\ln c_0-\mu-a\sigma^2}{\sigma}\right)\right].
\end{array}
\]
This, in turn, determines uniquely the value of $k$ since
$T_+(1)=1$.

So, in this example we have constructed
a class of distortions $T_+$ parameterized by
$z_0=F^{-1}(c_0)\in (0,1), a<0$ and $b\in (0,1)$.
These distortions are reversed  S-shaped, and satisfy the monotonicity condition
in Assumption \ref{Fdec}.
}
\end{exam}


{\rm The expressions of $H(\cdot)$ given in (\ref{dis1}) and (\ref{dis2})
show that the distortion $T_+(\cdot)$ in effect distorts
the distribution of $\rho$, a lognormal random variable, into one
having lognormal components, albeit with enlarged means and rescaled values.
On the other hand,
as stipulated in Tversky and Kahneman (1992), a probability
distortion on gain usually satisfies
$T_+'(0)=T_+'(1)=+\infty$, reflecting the observation that
there are most significant distortions
on very small and very large probabilities.
It turns out that the distortion functions constructed in the preceding example do
indeed satisfy $T_+'(0)=T_+'(1)=+\infty$. To see this,
notice $T_+'(z)=T_+'(F(x))=H'(x)/F'(x)=kx^{j(x)}$ or $\tilde kx^{j(x)}$. Hence,
when $z\rightarrow 0$, $x\rightarrow 0$, and $T_+'(z)\rightarrow +\infty$.
On the other hand, when $z\rightarrow 1$, $x\rightarrow +\infty$, and $T_+'(z)\rightarrow +\infty$. }

\section{Negative Part Problem}
Now we turn to the negative part problem (\ref{-part}), which is a Choquet
{\it minimization} problem.
Such a problem in a more general setting is solved thoroughly in Appendix
\ref{nppsubsection}; so we need only to apply the results there to (\ref{-part}). Notice, though,
(\ref{-part}) has a constraint that a feasible solution
must be almost surely bounded from above. The reason we do not include
this constraint explicitly
into the general problem (\ref{minusutility}) is that, under a mild
condition, any optimal solution to (\ref{minusutility}) is {\it automatically}
almost surely bounded from above; see Proposition \ref{minv} and the comments
right after it.

Similarly with the
positive part problem, for a
given $A=\{\omega: \rho\le c\}$ with $\underline{\rho}\leq c<\bar\rho $
(the case when $c=\bar\rho$ is trivial), we define $T_{A^C}(x):=\frac{T_-(xP(A^C))}{T_-(P(A^C))}$.
Then $T_{A^C}(\cdot)$ is a strictly increasing, differentiable
function from $[0,1]$ to $[0,1]$ with
$T_{A^C}(0)=0, T_{A^C}(1)=1$. Moreover, for any feasible solution $X$ of
(\ref{-part}) and  any $y\ge0$,
$$T_-(P\{u_-(X)>y\})=T_-\left(P\{u_-(X)>y|A^C\}P(A^C)\right)=
T_-(P(A^C))T_{A^C}(P\{u_-(X)>y|A^C\}).$$
Define the probability measure $P_{A^C}(\cdot)=P(\cdot|A^C)$.
Then Problem (\ref{-part}), taken in the
probability space $\left(\Omega\cap A^C, \cF\cap A^C, P_{A^C}\right)$,
 is equivalent to
\begin{equation}\label{-part2}
\begin{array}{ll}
\mbox{\rm Minimize}& V_-(Y)=T_-(P(A^C))\int_0^{+\infty} T_{A^C}(P_{A^C}\{u_-(Y)>y\})dy\\
\mbox{\rm subject to} & E_{A^C}[\rho Y]=(x_+-x_0)/P(A^C), \;\; Y\ge 0, \;\;
Y \mbox{ is bounded }\as.
\end{array}
\end{equation}
This is a special case of (\ref{minusutility}) in Appendix \ref{nppsubsection}.
\begin{theo}\label{os-p2}
Assume that $u_-(\cdot)$ is strictly concave at $0$.
Given $A:=\{\omega: \rho\le c\}$ with
$\underline{\rho}\leq c\le \bar\rho$, and $x_+\ge x_0^+$.
\begin{itemize}
\item[{\rm (i)}] If $c=\bar\rho$ and $x_+=x_0$, then the optimal solution of (\ref{-part}) is $X^*=0$ and $v_-(c, x_+)=0$.
\item[{\rm (ii)}] If $c=\bar\rho$ and $x_+\neq x_0$,
then
there is no feasible solution to (\ref{-part}) and $v_-(c, x_+)=+\infty$.
\item[{\rm (iii)}] If $\underline{\rho}\leq c<\bar\rho$,
then $v_-(c, x_+)=\inf_{\bar c\in[c,\bar\rho)} u_-\left(\frac{x_+-x_0}{E[\rho\id_{\rho>\bar c}]}\right)T_-\left(1-F(\bar c)\right)$.
Moreover,
Problem (\ref{-part}) with parameters $(A,x_+)$ admits an optimal
solution $X^*$ if and only if the following minimization problem
\begin{equation}\label{mincr}
\min_{\bar c\in[c,\bar\rho)} u_-\left(\frac{x_+-x_0}{E[\rho\id_{\rho>\bar c}]}\right)T_-\left(1-F(\bar c)\right)
\end{equation}
admits an optimal solution $\bar c^*$, in which case
$X^*=\frac{x_+-x_0}{E[\rho\id_{\rho>\bar c^*}]} \id_{\rho>\bar c^*}$, \as.
\end{itemize}
\end{theo}

{\sc Proof}:
Cases (i) and (ii) are trivial. On the other hand,
given Theorem \ref{-general}, and noticing that $X^*=\frac{x_+-x_0}{E[\rho\id_{\rho>\bar c^*}]} \id_{\bar c^*\le\rho<c}$ is automatically
bounded, we can prove (iii)
similarly to that for Theorem \ref{os+p2}. \eof

\section{Proof of Main Results}\label{proof}
Now that we have solved the problems of the positive and negative parts
in Step 1,
we are ready to
solve our ultimate Problem (\ref{btps}) via
the optimization Problem (\ref{step2}) or equivalently,
Problem (\ref{step2c}), in Step 2, and hence prove the main results contained in
Theorem \ref{2stepeq3}.

Recall problem (\ref{simstep2}) formulated earlier.
The following lemma is straightforward by Theorem \ref{os-p2} and the convention
(\ref{convention}).

\begin{lemma}\label{u->v-}
For any feasible pair $(c, x_+)$ for Problem (\ref{step2c}), $u_-\left(\frac{x_+-x_0}{E[\rho\id_{\rho>c}]}\right)T_-(1-F(c))\ge v_-(c,x_+)$.
\end{lemma}

\begin{prop}\label{equivofut-v-}
Problems (\ref{step2c}) and (\ref{simstep2})
have the same supremum values.
\end{prop}
{\sc Proof:}
Denote by $\alpha$ and $\beta$ the supremum values of (\ref{step2c}) and (\ref{simstep2}) respectively.
By Lemma \ref{u->v-}, $\alpha\ge \beta$.
Conversely, we prove $\alpha\le \beta$.  First we assume that $\alpha<+\infty$.
For any $\epsilon>0$, there exists
$(c, x_+)$ feasible for (\ref{step2c}) such that
$v_+(c, x_+)-v_-(c, x_+)\ge \alpha-\epsilon$, and there exists $\bar c\in[c,\bar\rho]$ such that
$u_-(\frac{x_+-x_0}{E[\rho\id_{\rho>\bar c}]})T_-(1-F(\bar c))\le v_-(c, x_+)+\epsilon$.
Therefore
\[\begin{array}{rl}
v_+(\bar c, x_+)-u_-\left(\frac{x_+-x_0}{E[\rho\id_{\rho>\bar c}]}\right)T_-(1-F(\bar c))
&\ge v_+(\bar c, x_+)-v_-(c, x_+)-\epsilon\\
&\ge v_+(c, x_+)-v_-(c,x_+)-\epsilon\\
&\ge \alpha-2\epsilon.
\end{array}
\]
Letting $\epsilon\rightarrow 0$, we conclude $\alpha\le \beta$.

Next, if $\alpha=+\infty$, then for any $M\in \R$, there exists a feasible pair $(c, x_+)$ such that
$v_+(c, x_+)-v_-(c, x_+)\ge M$, and there is $\bar c\ge c$ with
$u_-(\frac{x_+-x_0}{E[\rho\id_{\rho>\bar c}]})T_-(1-F(\bar c))\le v_-(c, x_+)+M/2$.
Thus $v_+(\bar c, x_+)-u_-(\frac{x_+-x_0}{E[\rho\id_{\rho>\bar c}]})T_-(1-F(\bar c))\ge v_+(c, x_+)-v_-(c, x_+)-M/2\ge M/2$, which
implies that $\beta=+\infty$.
\eof

{\sc Proof of Theorem \ref{2stepeq3}:}
(i) If $X^*$ is optimal for (\ref{btps}), then by Theorem \ref{2stepeq2}
$(c^*, x_+^*)$ is optimal for (\ref{step2c})
and
  $(X^*)^+\id_{\rho\le c^*}$ and $(X^*)^-\id_{\rho>c^*}$ are respectively
optimal for Problems
(\ref{+part}) and (\ref{-part}) with parameters $(\{\omega: \rho\le c^*\},x_+^*)$.
We now show that with $(c,x_+)=(c^*, x_+^*)$ the minimum
in (\ref{mincr}) is achieved at $\bar c=c^*$, namely,
\begin{equation}\label{mid1}
 v_-(c^*, x_+^*)=u_-(\frac{x_+^*-x_0}{E[\rho\id_{\rho> c^*}]})T_-(1-F( c^*)).
\end{equation}
To this end, we first assume that $x_+^*=0$ (hence
$X^*=0\;\as$ and $c^*=\bar \rho$).
Then $x_0\leq x_+^*=0$.
If $x_0=0$, then (\ref{mid1}) is trivial. If $x_0<0$,
Theorem \ref{os-p2} yields that
$(X^*)^-\id_{\rho>c^*}$ has the following representation
\begin{equation}\label{repres}
(X^*)^-\id_{\rho>c^*}=\frac{x_+^*-x_0}{E[\rho\id_{\rho>\bar c^*}]} \id_{\rho>\bar c^*},\;\;\as.
\end{equation}
Recall $X^*<0$ on $\rho>c^*$, and $\frac{x_+^*-x_0}{E[\rho\id_{\rho>\bar c^*}]}>0$; so (\ref{repres}) implies
$c^*=\bar c^*$, and hence (\ref{mid1}), in view of Theorem \ref{os-p2}.

Next, if $x_+^*>0$,
then by Proposition \ref{strictincofV+}, 
we have
$v_+(\bar c, x_+^*)>v_+(c^*, x_+^*)$ for any $\bar c> c^*$
with $P\{c^*<\rho\le \bar c\}>0$.
If (\ref{mid1}) is not true, then it follows from
Theorem \ref{os-p2} that there exists $\bar c>c^*$ with $P\{c^*<\rho\le \bar c\}>0$ such that $v_-(c^*, x_+^*)=u_-(\frac{x_+^*-x_0}{E[\rho\id_{\rho> \bar c}]})T_-(1-F(\bar c))$.
Consequently,
\[ \begin{array}{rl}
v_+(\bar c, x_+^*)-v_-(\bar c, x_+^*)\geq &
v_+(\bar c, x_+^*)-u_-(\frac{x_+^*-x_0}{E[\rho\id_{\rho> \bar c}]})T_-(1-F(\bar c))\\
>&v_+(c^*, x_+^*)-v_-(c^*, x_+^*),
\end{array}
\]
violating the conclusion that $(c^*, x_+^*)$ is optimal for (\ref{step2c}).

Now, for any $(c,x_+)$ feasible for
(\ref{simstep2}),
\[\begin{array}{rl}
&v_+(c,x_+)-u_-(\frac{x_+-x_0}{E[\rho\id_{\rho> c}]})T_-(1-F(c))\\
\leq & v_+(c,x_+)-v_-(c,x_+)\\
\leq & v_+(c^*, x_+^*)-v_-(c^*, x_+^*)\\
=& v_+(c^*, x_+^*)-u_-(\frac{x_+^*-x_0}{E[\rho\id_{\rho> c^*}]})T_-(1-F(c^*)),
\end{array}
\]
implying that $(c^*, x_+^*)$ is optimal
for (\ref{simstep2}).
The other conclusions are straightforward.

(ii) Since $(c^*, x_+^*)$ is optimal for (\ref{simstep2}), we have
\[\begin{array}{rl}
v_+(c^*, x_+^*)-v_-(c^*, x_+^*)
\geq & v_+(c^*, x_+^*)-u_-(\frac{x_+^*-x_0}{E[\rho\id_{\rho> c^*}]})T_-(1-F(c^*))\\
=&\sup\left[v_+(c, x_+)-u_-(\frac{x_+-x_0}{E[\rho\id_{\rho> c}]})T_-(1-F(c))\right]\\
=&\sup\left[v_+(c, x_+)-v_-(c, x_+)\right],
\end{array}
\]
where the supremum is over the feasible region of (\ref{simstep2}). This
implies that $(c^*, x_+^*)$ is optimal for (\ref{step2c}) and
the inequality above is in fact an equality, resulting in
\[ v_-(c^*, x_+^*)=u_-(\frac{x_+^*-x_0}{E[\rho\id_{\rho> c^*}]})T_-(1-F(c^*)).
\]
The above in turn indicates, thanks to Theorem \ref{os-p2}, that
$X_-^*:=\frac{x_+^*-x_0}{E[\rho\id_{\rho>c^*}]}\id_{\rho>c^*}$ is optimal
for (\ref{-part}) with parameters
$(\{\omega: \rho\le c^*\},x_+^*)$. The desired result then follows from
Theorem \ref{2stepeq2}.
\eof

Other claims in Section 4 on more explicit conclusions under Assumption \ref{Fdec}
are straightforward by virtue of Theorem \ref{os+p2}.

\section{ An Example with Two-Piece CRRA Utility Functions}
In this section we solve a concrete (and {\it very} involved) example
to demonstrate the general results obtained in previous section
as well as the algorithm presented.
The example showcases all the possibilities associated with
our behavioral portfolio selection model (\ref{btps0}), namely,
a model could be ill-posed, or well-posed yet optimal solution
not  attainable, or well-posed and optimal solution obtainable.
When the optimal solutions do exist, we are able to derive explicit
terminal payoffs for most of the cases.

In the example, we let $\rho$ follow the lognormal distribution, i.e.,
$\ln \rho\sim N(\mu, \sigma)$ with $\sigma>0$,
and the utility functions be CRRA ({\it constant relative
risk aversion}), i.e.,
$u_+(x)=x^{\alpha}, u_-(x)=k_-x^{\alpha}$, $x\ge0$,
with $k_->0$ and $0<\alpha<1$.
(Recall the overall utility function -- or value function in
the terminology of Tversky and Kahneman -- is an S-shaped function.)
These functions, also taken in Tversky and Kahneman (1992)
with $\alpha=0.88$ and $k_-=2.25$,
clearly
satisfy Assumption \ref{uassump}.
We do not spell out (neither do we need)
the explicit forms of the distortions $T_+(\cdot)$ and
$T_-(\cdot)$ so long as they satisfy Assumption \ref{Tassump}.
In addition, we assume that Assumption \ref{Fdec} holds,
which is imposed on $T_+(\cdot)$.
An example of such $T_+(\cdot)$ was presented in Example \ref{T+}.

Clearly, $u'_+(x)=\alpha x^{\alpha-1}, (u'_+)^{-1}(y)=(y/\alpha)^{1/(\alpha-1)},
u_+((u_+')^{-1}(y))=(y/\alpha)^{\alpha/(\alpha-1)}$, $\underline{\rho}=0$,
$\bar\rho=+\infty$, and
$F(x)=N\left((\ln x-\mu)/\sigma\right)$.



Under this setting, we first want to solve the positive part problem (\ref{+part})
with given $(c, x_+)$, where $0\leq c\leq +\infty$ and $x_+\geq x_0^+$.
The case that $c=0$ is trivial, where necessarily $x_+=0$ in order to
have a feasible problem, and
$v_+(c,x_+)=0$. So let $c\in(0,+\infty]$.
The optimal solution to (\ref{+part}) in this case is
$$X^*_+(c, x_+)=(u_+')^{-1}\left(\frac{\lambda(c, x_+) \rho}{T_+'(F(\rho))}\right)\id_{\rho\le c}
=\left(\frac{\lambda(c, x_+) \rho}{\alpha T_+'(F(\rho))}\right)^{1/(\alpha-1)}\id_{\rho\le c}.
$$
To determine $\lambda(c, x_+)$, denote
\[\varphi(c):=E\left[\left(\frac{T_+'(F(\rho))}{\rho}\right)^{1/(1-\alpha)}\rho\id_{\rho\le c}\right]>0,\;\;0< c\leq +\infty.
\]
Then the constraint $x_+=E[\rho X_+^*(c, x_+)]=\varphi(c)\left(\frac{\lambda(c, x_+)}{\alpha}\right)^{1/(\alpha-1)}$ gives
\[ \lambda(c,x_+)=\alpha\left(\frac{x_+}{\varphi(c)}\right)^{\alpha-1},\;\;0<c\leq +\infty,\;x_+\geq x_0^+.
\]
This in turn determines
\begin{equation}\label{X^*_1}
X^*_+(c, x_+)=\frac{x_+}{\varphi(c)}\left(\frac{T_+'(F(\rho))}{\rho}\right)^{1/(1-\alpha)}\id_{\rho\le c}, \;\;0<c\leq +\infty,\;x_+\geq x_0^+,
\end{equation}
and 
\begin{equation}\label{vcx}
\begin{array}{rcl}
v_+(c, x_+)
&=&\left(\frac{x_+}{\varphi(c)}\right)^\alpha E\left[\left(\frac{\rho}{T'_+(F(\rho))}\right)^{\alpha/(\alpha-1)-1}\rho\id_{\rho\le c}\right]\\
&=&\left(\frac{x_+}{\varphi(c)}\right)^\alpha \varphi(c)\\
&=&\varphi(c)^{1-\alpha}x_+^\alpha,\;\;0<c\leq +\infty,\;x_+\geq x_0^+.
\end{array}
\end{equation}
Set $\tilde\varphi(c)=\left\{\begin{array}{ll}
\varphi(c) & \mbox{ if } 0<c\leq +\infty,\\
0& \mbox{ if } c=0,
\end{array}\right.$  which is a non-decreasing function right continuous at 0.
Then Problem (\ref{simstep2}) specializes to
\begin{equation}\label{exr2opt}
\begin{array}{ll}
\mbox{\rm Maximize}& v(c, x_+)= \tilde\varphi(c)^{1-\alpha}x_+^\alpha -\frac{k_-T_-(1-F(c))}{(E[\rho\id_{\rho>c}])^\alpha}(x_+-x_0)^\alpha,\\
\mbox{\rm subject to} & \left\{\begin{array}{l}
0\le c\leq +\infty,\;\;x_+\ge x_0^+,\\
x_+=0 \mbox{ when } c=0,\;\;x_+=x_0 \mbox{ when } c=+\infty.
                         \end{array}\right.
\end{array}
\end{equation}
When $c>0$ ({\it ex}cluding $+\infty$) and $x_+\ge x_0^+$,
write
\[ v(c, x_+)= \varphi(c)^{1-\alpha}[x_+^\alpha -k(c)(x_+-x_0)^\alpha],
\]
where $k(c):=\frac{k_-T_-(1-F(c))}{\varphi(c)^{1-\alpha}(E[\rho\id_{\rho>c}])^\alpha}
>0, \;\;c>0$.

\medskip

We study the underlying portfolio selection problem in two cases, depending on
whether the initial wealth represents a gain or a loss.

\begin{theo}\label{a=b_x>0}
Assume that $x_0\geq 0$ and Assumption \ref{Fdec}
holds.
\begin{itemize}
\item[{\rm (i)}] If $\inf_{c> 0} k(c)\ge 1$, then
the optimal portfolio for Problem (\ref{btps0}) is
the replicating portfolio for the contingent claim
$$X^*=\frac{x_0}{\varphi(+\infty)}\left(\frac{T_+'(F(\rho))}{\rho}\right)^{1/(1-\alpha)}.$$
\item[{\rm (ii)}] If $\inf_{c>0} k(c)< 1$, then Problem (\ref{btps0})
is ill-posed. 
\end{itemize}
\end{theo}
{\sc Proof:}
Consider the problem $\max_{x\ge x_0} f(x)$
where $f(x)=x^\alpha-k(x-x_0)^\alpha$ and $k\geq0$ fixed.
Since $f'(x)=\alpha[x^{\alpha-1}-k(x-x_0)^{\alpha-1}]$,
we conclude that 1) if $k\ge 1$, then
$f'(x)\le 0\;\forall x\geq x_0$; therefore $x^*=x_0$ is optimal
with the
optimal value $x_0^\alpha$; and 2) if
$k<1$, then $f(x)=x^\alpha[1-k(1-x_0/x)^\alpha]\rightarrow +\infty$ as
$x\rightarrow +\infty$, implying that
$\sup_{x\geq x_0} f(x)=+\infty$.

(i) If $\inf_{c>0} k(c)\ge 1$, then
\[ \begin{array}{rl}
\sup_{c>0,x_+\ge x_0^+} v(c, x_+)
\equiv & \sup_{c> 0}\left[\varphi(c)^{1-\alpha}
\sup_{x_+\ge x_0}\left(x_+^\alpha-k(c)(x_+-x_0)^\alpha\right)\right]\\
=&\sup_{c>0}[\varphi(c)^{1-\alpha}x_0^\alpha]=\varphi(+\infty)^{1-\alpha}x_0^\alpha
\equiv v_+(+\infty,x_0)\geq0.
\end{array}
\]
However, when $c=0$ (and hence $x_+=0$) we have $v(c, x_+)=0$.
As a result
$(c^*,x_+^*)=(+\infty,x_0)$ is optimal to (\ref{exr2opt}).
Theorem \ref{2stepeq3} then applies to
conclude that $X^*\equiv X^*_+(+\infty,x_0)=\frac{x_0}{\varphi(+\infty)}\left(\frac{T_+'(F(\rho))}{\rho}\right)^{1/(1-\alpha)}$
solves (\ref{btps}).
Hence the optimal portfolio for (\ref{btps0}) is the one that replicates $X^*$.

(ii) If $\inf_{c>0} k(c)<1$, then there is $c_0>0$ such that $k(c_0)<1$.
In this case,
\[
\sup_{c>0,x_+\ge x_0^+} v(c, x_+)\geq j(c_0)^\alpha
\sup_{x_+\ge x_0}\left[x_+^\alpha-k(c_0)(x_+-x_0)^\alpha\right]=+\infty.
\]
The conclusion thus follows from
Propositions \ref{equivofut-v-} and \ref{ill=ill}.
.
\eof

\begin{theo}\label{x0s0}
Assume that $x_0< 0$ and Assumption \ref{Fdec} holds.
\begin{itemize}
\item [{\rm (i)}]
If $\inf_{c>0} k(c)> 1$, then Problem (\ref{btps0}) is well-posed. Moreover,
(\ref{btps0}) admits
an optimal portfolio
if and only if
\begin{equation}\label{extra}
{\rm argmin}_{c\geq0}\left[\left(\frac{k_-T_-(1-F(c))}{(E[\rho\id_{\rho>c}])^\alpha}\right)^{1/(1-\alpha)}-\tilde\varphi(c)\right]\neq \O.
\end{equation}
Furthermore, if $c^*>0$ is one of the minimizers
in (\ref{extra}), then the optimal portfolio is the one to replicate
\begin{equation}\label{xstar}
X^*=\frac{x^*_+}{\varphi(c^*)}\left(\frac{T_+'(F(\rho))}{\rho}\right)^{1/(1-\alpha)}\id_{\rho\le c^*}
     -\frac{x^*_+-x_0}{E[\rho\id_{\rho>c^*}]}\id_{\rho>c^*},
\end{equation}
where $x^*_+:=\frac{-x_0}{k(c^*)^{1/(1-\alpha)}-1}$; and
if $c^*=0$ is the unique minimizer in (\ref{extra}), then the unique optimal portfolio is the one to replicate $X^*=\frac{x_0}{E\rho}$.

\item [{\rm (ii)}] If $\inf_{c>0} k(c)=1$, then the supremum value of
Problem (\ref{btps0}) is $0$, which is however not achieved
by any admissible portfolio.
\item [{\rm (iii)}] If $\inf_{c>0} k(c)< 1$, then Problem (\ref{btps0}) is ill-posed.
\end{itemize}
\end{theo}

{\sc Proof:}
We first consider a general optimization problem $\max_{x\ge 0} f(x)$
where $f(x):=x^\alpha-k(x-x_0)^\alpha$ and $k\geq0$ fixed.
We solve it in the following three cases.
\begin{itemize}
\item [1)] If $k>1$, then $f'(x)=0$ has the only solution
$x^*=\frac{-x_0}{k^{1/(1-\alpha)}-1}>0$.
Since $f^\dpm(x)=\alpha(\alpha-1)[x^{\alpha-2}-k(x-x_0)^{\alpha-2}]<0\;\forall x>0$,
$x^*$ is the (only) maximum point with the maximum value
\[
  f(x^*)=(x^*)^\alpha[1-k(1-x_0/x)^\alpha]
=-(-x_0)^\alpha[k^{1/(1-\alpha)}-1]^{1-\alpha}.
\]

\item [2)] If $k=1$, then $f'(x)>0$ $\forall x> 0$. This means
that the supremum of $f(x)$ on $x\geq0$ is $\lim_{x\rightarrow +\infty}f(x)=0$;
yet this value is not achieved by any $x\geq0$.

\item [3)] If $k<1$, then $f(x)=x^\alpha[1-k(1-x_0/x)^\alpha]\rightarrow +\infty$ as
$x\rightarrow +\infty$, implying that
$\sup_{x\geq 0} f(x)=+\infty$.
\end{itemize}

We need to solve (\ref{exr2opt}) to obtain $(c^*,x_+^*)$. Since
$x_0<0$, $c=+\infty$ is infeasible;
so we restrict $c\in [0,+\infty)$. Care must be taken to
deal with the special solution $(c,x_+)=(0,0)$ with
$v(0,0)=-\frac{k_-}{(E[\rho])^\alpha}(-x_0)^\alpha$.

(i) If $\inf_{c>0} k(c)> 1$, then
\begin{equation}\label{sss}
\begin{array}{rl}
\sup_{c>0, x_+\ge x_0^+} v(c, x_+)\equiv & \sup_{c>0}\left[\varphi(c)^{1-\alpha}
\sup_{x_+\ge 0}\left(x_+^\alpha-k(c)(x_+-x_0)^\alpha\right)\right]\\
=&\sup_{c>0}\left[-(-x_0)^\alpha \varphi(c)^{1-\alpha}\left(k(c)^{1/(1-\alpha)}-1\right)^{1-\alpha}\right]\\
=&-(-x_0)^\alpha \left\{\inf_{c>0}\left[\left(\frac{k_-T_-(1-F(c))}{(E[\rho\id_{\rho>c}])^\alpha}\right)^{1/(1-\alpha)}-\varphi(c)\right]\right\}^{1-\alpha}<+\infty.
\end{array}
\end{equation}
This yields that (\ref{btps0}) is well-posed.
Now, {if $c^*>0$ achieve the infimum
of $\left[\left(\frac{k_-T_-(1-F(c))}{(E[\rho\id_{\rho>c}])^\alpha}\right)^{1/(1-\alpha)}-\tilde\varphi(c)\right]$
over $c\geq0$, then we have $\sup_{c>0, x_+\ge x_0^+} v(c, x_+)\geq -(-x_0)^\alpha\frac{k_-}{(E\rho)^\alpha}=v(0,0)$,
which means $c^*>0$ and $x^*_+=\frac{-x_0}{k(c^*)^{1/(1-\alpha)}-1}$
are optimal for (\ref{exr2opt}).
Theorem \ref{2stepeq3} then yields that the optimal portfolio
is the one that replicates $X^*$ given by (\ref{xstar}).

{If $c^*=0$ is the unique infimum
of $\left[\left(\frac{k_-T_-(1-F(c))}{(E[\rho\id_{\rho>c}])^\alpha}\right)^{1/(1-\alpha)}-\tilde\varphi(c)\right]$ over $c\geq0$,
then
\begin{eqnarray*}
\sup_{c>0, x_+\ge x_0^+} v(c, x_+)
&=&-(-x_0)^\alpha \left\{\inf_{c>0}\left[\left(\frac{k_-T_-(1-F(c))}{(E[\rho\id_{\rho>c}])^\alpha}\right)^{1/(1-\alpha)}-\varphi(c)\right]\right\}^{1-\alpha}\\
&<&-(-x_0)^\alpha \frac{k_-}{(E\rho)^\alpha}\equiv v(0,0).
\end{eqnarray*}
This implies that $(c^*,x_+^*)=(0,0)$ is uniquely optimal for
(\ref{exr2opt}), and the unique optimal solution for
(\ref{btps}) is
$X^*_+(c^*,x_+^*)\id_{\rho\leq c^*}-\frac{x_+^*-x_0}{E[\rho\id_{\rho>c^*}]}\id_{\rho>c^*}\equiv
\frac{x_0}{E\rho}$, for which the corresponding
replicating portfolio is the risk-free one.}

{If  the infimum
$\inf_{c\geq 0}\left[\left(\frac{k_-T_-(1-F(c))}{(E[\rho\id_{\rho>c}])^\alpha}\right)^{1/(1-\alpha)}-\tilde\varphi(c)\right]$
is not attainable, then\\
$\sup_{c>0, x_+\ge x_0^+}v(c,x_+)>-\frac{k_-}{(E[\rho])^\alpha}(-x_0)^\alpha
=v(0,0)$.
This means that $(0,0)$ is not optimal for (\ref{exr2opt}).
On the other hand, the optimality of (\ref{exr2opt}) is not achieved at
any $c>0$ and $x_+\ge0$ in view of (\ref{sss}).
It then follows from Theorem \ref{2stepeq3}
that (\ref{btps0}) admits no
optimal solution.}

(ii) Next consider the case when $\inf_{c>0}k(c)=1$.
If $k(c)>1$ for any $c>0$,
then
\begin{eqnarray*}
\sup_{c>0,x_+\ge x_0^+} v(c, x_+)
&=&-(-x_0)^\alpha \{\varphi(c)^{1-\alpha}[(\inf_{c>0}k(c))^{1/(1-\alpha)}-1]^{1-\alpha}\}=0.
\end{eqnarray*}
{Yet, for any $c>0, x_+\ge 0$,
$v(c, x_+)\le \max_{x_+\ge 0}v(c,x_+)
=-(-x_0)^\alpha [\varphi(c)^{1-\alpha}(k(c)^{1/(1-\alpha)}-1)^{1-\alpha}]
<0$. Also, $v(0,0)<0$. Therefore the optimal value is not attainable.
}

On the other hand, if there exists $c^*>0$ such that $k(c^*)=1$, then
$\sup_{c\geq0,x_+\ge x_0^+}v(c,x_+)\ge \sup_{x_+\ge 0}v(c^*, x_+)=0$.
However,
$v(c, x_+)=\varphi(c)^{1-\alpha}[x_+^{\alpha}-k(c)(x_+-x_0)^\alpha]\le \varphi(c)^{1-\alpha}[x_+^{\alpha}-(x_+-x_0)^\alpha]<0$ $\forall c>0,\;x_+\ge 0$.
Together with the fact that $v(0,0)<0$ we conclude that
$\sup_{c\geq0,x_+\ge x_0^+}v(c,x_+)=0$,
which is however not achieved.

(iii) If $\inf_{c\geq0}k(c)<1$, then there exists $c_0$ such that
$k(c_0)<1$. As a result,
$v(c_0,x_+)=\varphi(c_0)^{1-\alpha}[x_+^\alpha-k(c_0)(x_+-x_0)^\alpha]\rightarrow +\infty$ as $x_+\rightarrow +\infty$.
\eof

{\rm We see that the key features of
the
underlying behavioral portfolio selection problem critically depend on the
value $\inf_{c>0} k(c)$. Recall that $k(c)$, by its definition,
reflects in a precise way
the coordination among the utility functions, the
probability distortions, and the market (represented by $\rho$).
Let us elaborate on one particular point. In Tversky and Kahneman (1992),
the parameters are taken, based on extensive experiments,
to be $\alpha=0.88$, and $k_-=2.25>1$, the latter reflecting the
fact that losses loom larger than gains: the pain associated with
a loss is typically larger than the pleasure associated with an equivalent
gain\footnote{$k_-$ is the so-called {\it loss aversion coefficient}.}. Now, by the definition of $k(c)$ we see the larger
the loss aversion the more likely the underlying model is well-posed and
solvable. The economic intuition behind this is that with a larger 
loss aversion coefficient it is not optimal to allocate all the fund to 
stocks (because stocks are risky and prone to losses), and hence one needs to carefully balance the investment between risky and risk-free assets, leading to a meaningful model.
}

{\rm Another interesting observation is that
the optimal portfolios behave fundamentally different depending
on whether $x_0>0$ (Theorem \ref{a=b_x>0}) or $x_0<0$ (Theorem \ref{x0s0}).
Recall that the state $0$ here really means the reference point (e.g.,
the present value of a future liability that must be fulfilled); therefore
the two situations correspond to whether the investor starts with
a gain or loss situation.
If $x_0>0$, then the optimal strategy is simply to spend $x_0$ buying a
contingent claim that delivers a payoff in excess of the reference point,
reminiscent of a classical utility maximizing agent (although the allocation
to stocks is ``distorted'' due to the probability distortion).
If $x_0<0$, then the investor starts off a loss situation and needs to 
get ``out of the hole'' soonest possible. As a result, 
the optimal strategy is a gambling policy which
involves raising
additional capital to purchase a claim that delivers a higher payoff
in the case of a good state of the market and incurs a fixed loss in the case of a bad one.
Finally, if $x_0=0$, then
the optimal portfolio is {\it not} to invest in risky asset at all. Notice 
that $x_0=0$ corresponds to a natural psychological reference point -- 
the risk-free return -- for many people. 
This, nonetheless, does explain why most households do not invest in equities 
at all\footnote{A similar result is derived in Gomes (2005) for his portfolio selection model with loss averse
investors, albeit in the single-period setting without probability distortions.}.
}

\section{How Behavioral Criterion Affects Risky Allocation}\label{epp}

Along the line of the discussions at the end of the last section
we would like to investigate more on
how exactly the behavioral criterion would
affect the wealth allocation to risky assets.
This is best explained
through a very concrete example, where an optimal portfolio (not just
optimal terminal payoff) is explicitly available.
We consider a model with the power utility $u_+(x)=x^\alpha, u_-(x)=k_-x^\alpha$,
and all the market parameters (investment opportunity set)
are time-invariant: $r(\cdot)\equiv r, B(\cdot)=B, \sigma(t)=\sigma, \theta(\cdot)=\theta$.
In this case $\rho(t,T):=\rho(T)/\rho(t)$, given ${\cF}_t$, follows a lognormal distribution with parameter $(\mu_t,\sigma_t^2)$, where
\begin{equation}\label{log}
\mu_t:=-(r+\theta^2/2)(T-t),\;\;\sigma_t^2:=\theta^2(T-t).
\end{equation}
Furthermore, we set the distortion $T_+$ to be the one
in Example \ref{T+} with $j(x)=:a\id_{0<x\le c_0}+b\id_{x>c_0}$, where $c_0>0$, $a<0$ and $0<b<1$.

We now derive in closed-form the optimal portfolio under the setting of
Theorem \ref{a=b_x>0}-(i), i.e., $x_0\geq0$ and $\inf_{c> 0} k(c)\ge 1$.
(Other cases can also be done, which are left to interested readers.)

\begin{theo}\label{examp}
Under the assumption of Theorem \ref{a=b_x>0}-(i),
the optimal wealth-portfolio pair $(x^*(\cdot), \pi^*(\cdot))$ for
Problem (\ref{btps0}) is
{\small \begin{eqnarray*}
     x^*(t)&=&\frac{x_0}{\gamma }[x^1(t)+c_0^{(a-b)/(1-\alpha)}x^2(t)],\\
 \pi^*(t)&=&\frac{x_0}{\gamma }
\left[\frac{(1-a)x^1(t)+c_0^{(a-b)/(1-\alpha)}(1-b)x^2(t)}{1-\alpha}\right]
(\sigma\sigma')^{-1}B,
 \end{eqnarray*}}
where $\psi(y):=(2\pi)^{-1/2}e^{-y^2/2}$ is the
density function of
a standard normal distribution, and
{
\[\begin{array}{l}
x^1(t):=\frac{\rho(t)^{(a-1)/(1-\alpha)}}{\sigma_t}\int_0^{c_0/\rho(t)}y^{(a-1)/(1-\alpha)}
          \psi\left(\frac{\ln y-\mu_t}{\sigma_t}\right)dy
\equiv \frac{1}{\sigma_t\rho(t)}\int_0^{c_0}y^{(a-1)/(1-\alpha)}
          \psi\left(\frac{\ln y-\mu_t-\ln\rho(t)}{\sigma_t}\right)dy,\\
\\
x^2(t):=\frac{\rho(t)^{(b-1)/(1-\alpha)}}{\sigma_t}\int_{c_0/\rho(t)}^{+\infty}y^{(b-1)/(1-\alpha)}
          \psi\left(\frac{\ln y-\mu_t}{\sigma_t}\right)dy
\equiv \frac{1}{\sigma_t\rho(t)}\int_{c_0}^{+\infty}y^{(b-1)/(1-\alpha)}
          \psi\left(\frac{\ln y-\mu_t-\ln\rho(t)}{\sigma_t}\right)dy,\\
\\
\gamma:=E\left[\rho^{(a-\alpha)/(1-\alpha)}\id_{\rho\le c_0}+c_0^{(a-b)/(1-\alpha)}\rho^{(b-\alpha)/(1-\alpha)}\id_{\rho> c_0}\right].
\end{array}
\]}
\end{theo}

{\sc Proof:}
It follows from (\ref{Ha}) and (\ref{Hb}) that
\[
\frac{T_+'(F(\rho))}{\rho}\equiv\frac{H'(\rho)}{\rho F'(\rho)}
=k\rho^{a-1}\id_{\rho\le c_0}+kc_0^{a-b}\rho^{b-1}\id_{\rho>c_0},
\]
where $k^{-1}=e^{a\mu_0+a^2\sigma_0^2/2}N(\frac{\ln c_0-\mu_0-a\sigma_0^2}{\sigma_0})
  +c_0^{a-b}e^{b\mu_0+b^2\sigma_0^2/2}[1-N(\frac{\ln c_0-\mu_0-a\sigma_0^2}{\sigma_0})]$.
Hence
\[ \varphi(+\infty):=E\left[\left(\frac{T_+'(F(\rho))}{\rho}\right)^{\frac{1}{1-\alpha}}\rho\right]
=k^{\frac{1}{1-\alpha}}E\left[\rho^{\frac{a-\alpha}{1-\alpha}}\id_{\rho\le c_0}
+c_0^{\frac{a-b}{1-\alpha}}\rho^{\frac{b-\alpha}{1-\alpha}}\id_{\rho> c_0}\right]=k^{\frac{1}{1-\alpha}}\gamma.
\]
Appealing to Theorem \ref{a=b_x>0}-(i) the optimal portfolio is the replicating portfolio for the claim
\[ X^*=\frac{x_0}{\gamma }[\rho^{(a-1)/(1-\alpha)}\id_{\rho\le c_0}+c_0^{(a-b)/(1-\alpha)}\rho^{(b-1)/(1-\alpha)}\id_{\rho> c_0}].
\]
Let $(x^1(\cdot), \pi^1(\cdot))$ replicate $\rho^{(a-1)/(1-\alpha)}\id_{\rho\le c_0}$ and
$(x^2(\cdot), \pi^2(\cdot))$ replicate $\rho^{(b-1)/(1-\alpha)}\id_{\rho> c_0}$.
Then the results in
Appendix B yield
 \begin{eqnarray*}
  x^1(t)&=&\frac{\rho(t)^{(a-1)/(1-\alpha)}}{\sigma_t}\int_0^{c_0/\rho(t)}y^{(a-1)/(1-\alpha)}
          \psi\left(\frac{\ln y-\mu_t}{\sigma_t}\right)dy,\\
   \pi^1(t)&=&-\left[\frac{a-1}{1-\alpha}x^1(t)-\frac{1}{\sigma_t\rho(t)}
  c_0^{(a-\alpha)/(1-\alpha)}\psi\left(\frac{\ln c_0-\mu_t-\ln \rho(t)}{\sigma_t}\right)
  \right](\sigma\sigma')^{-1}B,\\
    x^2(t)&=&\frac{\rho(t)^{(b-1)/(1-\alpha)}}{\sigma_t}\int_{c_0/\rho(t)}^{+\infty}y^{(b-1)/(1-\alpha)}
          \psi\left(\frac{\ln y-\mu_t}{\sigma_t}\right)dy,\\
   \pi^2(t)&=&-\left[\frac{b-1}{1-\alpha}x^2(t)+\frac{1}{\sigma_t\rho(t)}
  c_0^{(b-\alpha)/(1-\alpha)}\psi\left(\frac{\ln c_0-\mu_t-\ln \rho(t)}{\sigma_t}\right)
  \right](\sigma\sigma')^{-1}B.
 \end{eqnarray*}
Combining these two portfolios linearly we obtain the desired result.
\eof

Now consider the case when $c_0=1$ for ease of exposition.
In this case the optimal portfolio can be simplified to be
\begin{eqnarray*}
\pi^*(t)&=&\frac{x_0}{\gamma }\left[\frac{(1-a)x^1(t)+(1-b)x^2(t)}{1-\alpha}\right](\sigma\sigma')^{-1}B\\
&=&\frac{(1-a)x^1(t)+(1-b)x^2(t)}{x^1(t)+x^2(t)}\frac{x^*(t)}{1-\alpha}(\sigma\sigma')^{-1}B\\
&=&\left(1-\frac{ax^1(t)+bx^2(t)}{x^1(t)+x^2(t)}\right)\frac{x^*(t)}{1-\alpha}(\sigma\sigma')^{-1}B,
\end{eqnarray*}
or the optimal ratio in risky assets is
\begin{equation}\label{ratio}
\frac{\pi^*(t)}{x^*(t)}=(1-\alpha)^{-1}\left(1-\frac{ax^1(t)+bx^2(t)}{x^1(t)+x^2(t)}\right)(\sigma\sigma')^{-1}B.
\end{equation}

Recall that in the conventional expected utility model
with the utility function $u(x)=x^\alpha$ and without distortion,
the optimal ratio in risky assets is
\begin{equation}\label{ratio2}
\frac{\hat\pi(t)}{\hat x(t)}=(1-\alpha)^{-1}(\sigma\sigma')^{-1}B.
\end{equation}

So when
\[\frac{b}{-a}>\frac{x^1(t)}{x^2(t)}
=\frac{\int_0^1y^{(a-1)/(1-\alpha)}\psi\left(\frac{\ln y-\mu_t-\ln\rho(t)}{\sigma_t}\right)dy}
      {\int_1^{+\infty}y^{(b-1)/(1-\alpha)}\psi\left(\frac{\ln y-\mu_t-\ln\rho(t)}{\sigma_t}\right)dy},
\]
      the investor {\it underweights}
the risky assets in her portfolio compared with the one
dictated by the conventional utility model, and {\it vice versa}.

\section{Concluding Remarks}

In this paper, we introduce, for the first time in literature to our best
knowledge, a general continuous-time
portfolio selection model within the framework of the cumulative prospect
theory, so as to account for human psychology and emotions
in investment activities.
 The model features inherent difficulties, including
non-convex/concave and non-smooth (overall) utility functions and
probability distortions. Even the well-posedness of such a model becomes
more an exception than a rule: we demonstrate that
a well-posed model calls for
a careful coordination among the underlying market, the utility function, and
the probability distortions.
We then develop an approach to solving the model
thoroughly. The approach is
largely different from the existing ones employed
in the conventional dynamic asset allocation models.
Notwithstanding the complexity of the approach, the final solution
turns out to be simply structured: the optimal terminal payoff
is related to certain {binary options} characterized by a {single} number, and
the optimal strategy is an aggressive gambling policy betting on good states of the
market.
Finally, we apply the general results to
a specific case with a two-piece CRRA utility function, and show how the behavioral criterion will change the risky allocation.

The equity premium puzzle [Mehra and Prescott (1985)]
refers to the phenomenon that observed average
annual returns on stocks over the past century are higher by large
margin (approximately 6
percentage points) than returns on government bonds, whereas standard asset allocation theories (such as that based on the utility model) predict that the difference in returns between these two investments
should be much smaller.
Benartzi and
Thaler (1995) proposed an explanation for the puzzle using prospect theory
(in single period and without probability distortion).
In Section \ref{epp} we demonstrate that the investor would indeed
underweight stocks in her portfolio under certain conditions.
We are not claiming that we have provided a satisfactory explanation to the equity premium puzzle in the continuous time setting; but we do hope that
the research along the line will shed lights on eventually solving the puzzle.

It should be emphasized again that the agent under study in this paper is a ``small
investor'' in that his behavior will not affect the market. Hence we can
still comfortably assume some market properties, such as the absence of
arbitrage and the market completeness, as usually imposed for the conventional
utility model. (It remains an interesting problem to
study a behavioral model in an incomplete market.)
It is certainly a
fascinating and challenging problem to study how the overall market
might be changed by the joint behaviors of investors; e.g., a ``behavioral''
capital asset pricing model.

Let us also mention about an on-going work [He and Zhou (2007)]
on behavioral portfolio choice in single period, featuring both S-shaped utilities and 
probability distortions. Perversely, the single-period model is equally difficult, and calls for a technique quite different from its continuous-time counterpart to tackle. Only some special cases have been solved, which are used to study the equity premium puzzle more closely.

To conclude, this work is meant to be initiating and inspiring, rather than
exhaustive and conclusive, for the research on intertemporal behavioral
portfolio allocation.

\bigskip


\newpage

{\small
\noindent{\Large \bf Appendix}

\appendix

\section{An Inequality}

\begin{lemma}\label{inequality}
Let $f$: $\R^+\mapsto \R^+$  be a non-decreasing function
with $f(0)=0$. Then
  $$xy\le \int_0^xf^{-1}(t)dt+\int_0^yf(t)dt\;\;\forall x\ge 0, y\ge 0,$$
  and the equality holds if and only if $f(y-)\le x\le f(y+)$.
\end{lemma}
\pf By interpreting the integrations involved as the appropriate areas,
we have
$$\int_0^yf(t)dt=yf(y)-\int_0^{f(y)}f^{-1}(t)dt.$$
Define $g(x,y):=\int_0^xf^{-1}(t)dt+\int_0^yf(t)dt-xy$. Then
\[
g(x,y)=y(f(y)-x)+\int_{f(y)}^xf^{-1}(t)dt=\int_{f(y)}^x(f^{-1}(t)-y)dt.
\]

We now consider all the possible cases. First, if $x<f(y-)$, then
$f^{-1}(t)\le y$ $\forall t<f(y)$. Therefore $g(x,y)=\int_x^{f(y)}(y-f^{-1}(t))dt\ge 0$.
Moreover, in this case there exists $z>x$ such that $z<f(y-)$, which
implies $y>f^{-1}(z)$ (otherwise $f(y-\epsilon)<z$ $\forall \epsilon>0$,
leading to $z\ge f(y-)$). The monotonicity of $f^{-1}$
yields
$y>f^{-1}(t)$ for any $t\le z$. Hence
$g(x,y)=\int_x^{f(y)}(y-f^{-1}(t))dt\ge \int_x^z(y-f^{-1}(t))dt>0$.

Next consider the case when $x\in [f(y-), f(y)]$.
Since $f^{-1}(t)\le y$ $\forall t<f(y)$, and
$f^{-1}(t)\ge y$ $\forall t>x\ge f(y-)$, we have
$g(x,y)=\int_x^{f(y)}(y-f^{-1}(t))dt=0$.

Symmetrically, we can prove that $g(x,y)=0$ when $x\in [f(y), f(y+)]$, and
$g(x,y)>0$ when $x>f(y+)$. The proof is complete.
\eof

\medskip

\section{Two Auxiliary Optimization Problems}

In this subsection we solve two auxiliary optimization problems, which play a key role in
simplifying the behavioral portfolio selection model.

Let $Y$ be a given {\it strictly} positive random variable on
 $(\Omega, \cF, P)$ with the
probability distribution function $F(\cdot)$. Let $G(\cdot)$ be
another given distribution function with $G(0)=0$.
Consider the following two optimization problems:
\begin{equation}\label{maxxy-F}
  \begin{array}{ll}
  \mbox{\rm Maximize}& E[XY]\\
  \mbox{\rm subject to} & P(X\le x)=G(x)\;\; \forall x\in \R,
  \end{array}
\end{equation}
and
\begin{equation}\label{minxy-F}
  \begin{array}{ll}
\mbox{\rm Minimize}& E[XY]\\
\mbox{\rm subject to} & P(X\le x)=G(x)\;\; \forall x\in \R.
  \end{array}
\end{equation}

These are two highly non-convex optimization problems.

\begin{lemma}\label{mfunc}
\begin{itemize}
\item [{\rm (i)}] Let $h(\cdot)$ be a non-decreasing function. 
If $X$ and $h(Y)$ share the same distribution, then
$E[XY]\le E[h(Y)Y]$ while the equality holds
if and only if $X\in [h(Y-),h(Y+)]$ \as.
\item [{\rm (ii)}] Let $h(\cdot)$ be a non-increasing function. 
If $X$ and $h(Y)$ share the same distribution, then
$E[XY]\ge E[h(Y)Y]$ while the equality holds
if and only if $X\in [h(Y-),h(Y+)]$ \as
\end{itemize}
\end{lemma}
\pf (i) First assume $h(0)=0$. Employing Lemma \ref{inequality},
together with
the assumption that  $X$ and $h(Y)$ have the same distribution, we have
\begin{eqnarray*}
  E[XY]&\le&E[\int_0^Xh^{-1}(u)du]+E[\int_0^Yh(u)du]\\
&=&E[\int_0^{h(Y)}h^{-1}(u)du]+E[\int_0^Yh(u)du]=E[h(Y)Y],
\end{eqnarray*}
and the equality holds if and only if $X\in [h(Y-),h(Y+)]$ a.s..

For the general case when $h(0)\neq0$, define $\bar h(x):=h(x)-h(0)$. Then
$$E[XY]=E[(X-h(0))Y]+h(0)EY\le E[\bar h(Y)Y]+h(0)EY=E[h(Y)Y].$$

(ii) It is straightforward by applying the result in (i) to $-X$ and $-h(Y)$.
\eof

\begin{theo}  \label{comono}
Assume that $Y$ admits no atom.
\begin{itemize}
\item [{\rm (i)}] Define $X_1^*:=G^{-1}(F(Y))$. Then $E[X_1^*Y]\ge E[XY]$
for any feasible solution $X$ of Problem (\ref{maxxy-F}).
If in addition $E[X_1^*Y]<+\infty$, then $X_1^*$ is the unique (in the sense
of almost surely) optimal solution for (\ref{maxxy-F}).
\item [{\rm (ii)}] Define $X_2^*:=G^{-1}(1-F(Y))$. Then $E[X_2^*Y]\le E[XY]$ for any feasible solution $X$ of
Problem (\ref{minxy-F}). If in addition $E[X_2^*Y]<+\infty$,
then $X_2^*$ is the unique
optimal solution for (\ref{minxy-F}).
\end{itemize}
\end{theo}

\pf 
First of all note that $Z:=F(Y)$ follows uniform distribution on the
(open or closed) unit interval.

(i) Define $h_1(x):=G^{-1}(F(x))$. Then
$P\{h_1(Y)\le x\}=P\{Z\le G(x)\}=G(x)$, and $h_1(\cdot)$ is non-decreasing.
By Lemma \ref{mfunc},
$E[X_1^*Y]\ge E[XY]$
for any feasible solution $X$ of Problem (\ref{maxxy-F}), where
$X^*_1:=h_1(Y)$.
Furthermore,
if $E[X^*_1Y]<+\infty$, and
there is $X$ which is optimal for (\ref{maxxy-F}), then
$E[XY]=E[X^*_1Y]$. By Lemma \ref{mfunc}, $X\in [h_1(Y-),h_1(Y+)]$ a.s..
Since $h_1(\cdot)$ is non-decreasing,
its set of discontinuous points is at
most countable. However, $Y$ admits no atom; hence $h_1(Y-)=h_1(Y+)=h_1(Y)$, a.s.,
which implies that $X=h_1(Y)=X^*_1$, a.s..
Therefore we have proved that $X^*_1$ is the unique optimal solution for
(\ref{maxxy-F}).

(ii) Define $h_2(x):=G^{-1}(1-F(x))$. It is immediate that
$P\{h_2(Y)\le x\}=G(x)$, and $h_2(\cdot)$ is
non-increasing. Applying Lemma \ref{mfunc} and a similar argument
as in (i) we obtain the desired result.
\eof

The preceding theorem shows that the optimal solution to
(\ref{maxxy-F}) is comonotonic with $Y$, and that to (\ref{maxxy-F}) is anti-comonotonic with $Y$.

\medskip

\section{A Choquet Maximization Problem}\label{pppsubsection}
Consider a {general} utility maximization problem
involving the Choquet integral:
\begin{equation}\label{plusutility}
\begin{array}{ll}
\mbox{\rm Maximize}& V_1(X)=\int_0^{+\infty} T(P\{u(X)>y\})dy\\
\mbox{\rm subject to} & E[\xi X]=a, \;\; X\ge 0,
\end{array}
\end{equation}
where $\xi$ is a given strictly positive random variable,
with no atom and whose distribution function is $F_\xi(\cdot)$,
$a\ge0$,
$T: [0,1]\mapsto [0,1]$ is a strictly increasing, differentiable function
with $T(0)=0,\;T(1)=1$,
and $u(\cdot)$ is a strictly concave, strictly increasing, twice
differentiable  function
with $u(0)=0$, $u'(0)=+\infty$, $u'(+\infty)=0$.

The case $a=0$ is trivial, where $X^*=0$ is the only feasible, and hence optimal, solution.
So we assume $a>0$ in what follows.
The difficulty with (\ref{plusutility}) is that it is a {\it non-convex}
optimization problem with a {\it constraint}; thus the normal technique
like Lagrange multiplier does not apply directly. The approach we develop here is
to change the decision variable and
turn
the problem into a convex problem through a {\it series} of transformations.
To start with, we have the following lemma.
\begin{lemma}\label{maxv+-a}
If Problem (\ref{plusutility})
admits an optimal solution $X^*$
whose distribution function is $G(\cdot)$, then
$X^*=G^{-1}(1-F_\xi(\xi)),\;\;\as.$
\end{lemma}
{\sc Proof:} Since
$a>0$, $G(t)\not\equiv 1$. Denote $\bar X:=G^{-1}(1-F_\xi(\xi))$. Notice that $1-F_\xi(\xi)\sim U(0,1)$; thus
$\bar X$ has the same distribution as $X^*$ and
$E[\xi\bar X]>0$.

If $X^*=\bar X\;\;\as$ is not true, then it follows from the uniqueness
result in Theorem \ref{comono}
that
$E[\xi\bar X]<E[\xi X^*]=a$. Define $X_1:=k\bar X$, where $k:=a/E[\xi\bar X]>1$. Then $X_1$ is feasible for
(\ref{plusutility}), and $V_1(X_1)>V_1(\bar X)=V_1(X^*)$, which contradicts the optimality of $X^*$.
\eof

{Lemma \ref{maxv+-a} implies that an optimal solution to (\ref{plusutility}),
if it exists, must be anti-comonotonic with $\xi$.}

Denote $Z:=1-F_\xi(\xi)$. Then $Z$ follows $U(0,1)$,
and $\xi=F_\xi^{-1}(1-Z),\as$, thanks to $\xi$ being atomless.
Lemma \ref{maxv+-a} suggests that in order to solve (\ref{plusutility}) one needs
only to seek among random variables in the form $G^{-1}(Z)$, where $G$ is
the distribution function of a nonnegative
random variable [i.e., $G$ is non-decreasing, c\`adl\`ag, with $G(0-)=0,\;G(+\infty)=1$].
Motivated by this observation, we introduce
the following problem
\begin{equation}\label{maxvg-a}
  \begin{array}{ll}
\mbox{\rm Maximize}& v_1(G):=\int_0^{+\infty}T(P\{u(G^{-1}(Z))>t\})dt\\
\mbox{\rm subject to} & \left\{\begin{array}{l}
E[G^{-1}(Z)F_\xi^{-1}(1-Z)]=a,\\
G \mbox{ is the distribution function of a nonnegative random variable.}
\end{array}\right.
  \end{array}
\end{equation}

The following result, which is straightforward in view of Lemma \ref{maxv+-a},
 stipulates that Problem (\ref{maxvg-a}) is equivalent
to Problem (\ref{plusutility}).
\begin{prop}\label{disopt}
  If $G^*$ is optimal for (\ref{maxvg-a}), then
$X^*:=(G^*)^{-1}(Z)$
  is optimal for (\ref{plusutility}).
Conversely, if $X^*$ is optimal for
(\ref{plusutility}), then
its distribution function $G^*$ is optimal for (\ref{maxvg-a}) and $X^*=(G^*)^{-1}(Z),\;\as$.
\end{prop}

Now we turn to Problem (\ref{maxvg-a}). Denoting $\bar T(x):=T(1-x)$,
$x\in [0,1]$, and $\bar u:=\sup_{x\in\sR\!^+} u(x)$, we have
\begin{eqnarray*}
v_1(G)&=&\int_0^{\bar u}\bar T \left(P\{u(G^{-1}(Z))\le y\}\right)dy=
\int_0^{\bar u}\bar T \left(P\{Z\le G(u^{-1}(y))\}\right)dy\\
&=& \int_0^{\bar u}\bar T(G(u^{-1}(y)))dy=\int_0^1u(G^{-1}(\bar T^{-1}(t)))dt\\
&=&-\int_0^1u(G^{-1}(s))\bar T'(s)ds=\int_0^1u(G^{-1}(s))T'(1-s)ds\\
&=&E\left[u(G^{-1}(Z)) T'(1-Z)\right].
\end{eqnarray*}
Denoting
$$\Gamma:=\{g: [0,1)\mapsto \R^+ \mbox{ is non-decreasing, left continuous,
with } g(0)=0\},$$
and considering $g=G^{-1}$,
we can rewrite Problem (\ref{maxvg-a}) into
\begin{equation}\label{maxvg-a-rv}
\begin{array}{ll}
\mbox{\rm Maximize}& \bar v_1(g):=E[u(g(Z))T'(1-Z)]\\
\mbox{\rm subject to} & E[g(Z)F_\xi^{-1}(1-Z)]=a,\;\; g\in \Gamma.
\end{array}
\end{equation}

{\rm Some remarks on the set $\Gamma$ are in order.
Since any given $g\in \Gamma$ is left continuous, we can always
extend it to a map from $[0,1]$ to $\R^+\cup\{+\infty\}$ by setting
$g(1):=g(1-)$. It is easy to see that $g(1)<+\infty$ if and only if
the corresponding random variable $\eta$ (i.e.,
$\eta$ is such a random variable whose distribution function has
an inverse identical to
$g$) is almost surely bounded from
above.}

Since
$T'(\cdot)>0$ and $u(\cdot)$ is concave, the objective
functional of (\ref{maxvg-a-rv}) is now {\it concave} in $g$.
On the other hand, the constraint functional
$E[g(Z)F_\xi^{-1}(1-Z)]$ is linear in
$g$. Hence we can use the Lagrange method to remove this linear constraint
as follows.
For a given $\lambda\in\R$,
\begin{equation}\label{unc+part}
\begin{array}{ll}
\mbox{\rm Maximize}&
\tilde v_1^{\lambda}(g):=E\left[u(g(Z))T'(1-Z)-\lambda g(Z)F_\xi^{-1}(1-Z)\right]\\
\mbox{\rm subject to} & g\in \Gamma,
\end{array}
\end{equation}
and then determine $\lambda$ via the original linear constraint.

Although Problem (\ref{unc+part}) is a convex optimization problem
in $g$, it has an implicit constraint that $g$ be non-decreasing; hence
is very complex. Let us ignore this constraint for the moment.
For each fixed $z\in (0,1)$ we maximize
$u(g(z))T'(1-z)-\lambda g(z)F_\xi^{-1}(1-z)$ over $g(z)\in\R^+$.
The zero-derivative condition gives
$g(z)=(u')^{-1}(\lambda F_\xi^{-1}(1-z)/T'(1-z))$. Now, if
$F_\xi^{-1}(z)/T'(z)$ happens to be non-decreasing in $z\in(0,1]$,
then $g(z)$ is non-decreasing in $z\in[0,1)$ and, hence, it solves (\ref{unc+part}).
On the other hand, if $F_\xi^{-1}(z)/T'(z)$ is not non-decreasing, then
it remains an open problem to express {\it explicitly} the
optimal solution to (\ref{unc+part}).

\medskip

Denote $R_u(x):=-\frac{xu^\dpm(x)}{u'(x)},\;x>0$, which is
the {\it Arrow--Pratt index of relative risk aversion} of the utility function $u(\cdot)$.
\begin{prop}\label{lagequiv}
Assume that $F_\xi^{-1}(z)/T'(z)$ is non-decreasing in $z\in (0,1]$ and \\
$\liminf_{x\rightarrow +\infty}R_u(x)>0$.
Then the following claims are equivalent:
\begin{itemize}
\item [{\rm (i)}] Problem (\ref{maxvg-a-rv}) is well-posed for any $a>0$.
  \item [{\rm (ii)}] Problem (\ref{maxvg-a-rv}) admits a unique optimal solution for any $a>0$.
  \item [{\rm (iii)}] $E\left[u\left((u')^{-1}(\frac{\xi}{T'(F_\xi(\xi))})\right)T'(F_\xi(\xi))\right]<+\infty$.
\item [{\rm (iv)}] $E\left[u\left((u')^{-1}(\frac{\lambda\xi}{T'(F_\xi(\xi))})\right)T'(F_\xi(\xi))\right]<+\infty$ $\forall \lambda>0$.
 \end{itemize}
Furthermore, when one of the above (i)--(iv) holds, the optimal solution to
(\ref{maxvg-a-rv}) is
$$g^*(x)\equiv (G^*)^{-1}(x)=(u')^{-1}\left(\frac{\lambda F_\xi^{-1}(1-x)}{T'(1-x)}\right), \; x\in [0,1),$$
where $\lambda>0$ is the one satisfying
$E[(G^*)^{-1}(1-F_\xi(\xi))\xi]=a$.
\end{prop}

{\sc Proof:}
Since $T'(1-Z)>0$ and $E[T'(1-Z)]=\int_0^1T'(x)dx=T(1)-T(0)=1$,
we can define a new probability measure $\tilde P$
whose expectation $\tilde E(X):=E[ T'(1-Z)X]$.

Denote $\zeta:=\frac{F_\xi^{-1}(1-Z)}{T'(1-Z)}\equiv
\frac{\xi}{T'(F_\xi(\xi))}$. Then $\zeta>0\; \as$.
Rewrite Problem (\ref{maxvg-a-rv}) in terms of the probability measure $\tilde P$ as follows
\begin{equation}\label{maxvg-a-tilde}
\begin{array}{ll}
\mbox{\rm Maximize}& \bar v_1(g):=\tilde E[u(g(Z))]\\
\mbox{\rm subject to} & \tilde E[\zeta g(Z)] =a,\;\;
g\in \Gamma.
\end{array}
\end{equation}
By Jin, Xu and Zhou (2007, Theorem 6) and the fact that
$g^*(x)=(u')^{-1}\left(\frac{\lambda F_\xi^{-1}(1-x)}{T'(1-x)}\right)$ is automatically non-decreasing in $x$,
we get the desired result. \eof

We now summarize all the results above in the following theorem.
\begin{theo}\label{maxv}
Assume that $F_\xi^{-1}(z)/T'(z)$ is non-decreasing in $z\in (0,1]$ and
$\liminf_{x\rightarrow +\infty}R_u(x)>0$.
Define $X(\lambda):=(u')^{-1}\left(\frac{\lambda \xi}{T'(F_\xi(\xi))}\right)$
for $\lambda>0$.
If $V_1(X({1}))<+\infty$, then $X(\lambda)$ is an optimal solution for Problem (\ref{plusutility}),
where $\lambda$ is the one satisfying
$E[\xi X(\lambda)]=a$.
If $V_1(X({1}))=+\infty$, then 
Problem (\ref{plusutility}) is ill-posed.
\end{theo}

To conclude this subsection, we state
a necessary condition of optimality
for Problem (\ref{plusutility}), which
is useful in solving Problem (\ref{step2}) in Step 2.

\begin{lemma}\label{Ginv>0}
If $g$ is optimal for
(\ref{unc+part}), then either $g\equiv 0$ or $g(x)>0$ $\forall x>0$.
\end{lemma}

{\sc Proof}: Suppose $g\not\equiv 0$. We now show that
$g(x)>0$ $\forall x>0$. If not,
define $\delta:=\inf\{x>0: g(x)>0\}$.
Then $0<\delta<1$, and $g(x)=0$ $\forall x\in [0,\delta]$.

For any $y>0$, let $\epsilon(y):=\inf\{x>0: g(\delta+x)>y\}$
and
$$g_y(x):=\left\{\begin{array}{ll}
0,&x\in[0,\delta/2],\\ y,&x\in (\delta/2, \delta+\epsilon(y)],\\ g(x), & x\in(\delta+\epsilon(y),1).\end{array}\right.$$
Then
\begin{eqnarray*}
&&  E[u(g_y(Z))T'(1-Z)-\lambda g_y(Z)F_\xi^{-1}(1-Z)]-E[u(g(Z))T'(1-Z)-\lambda g(Z)F_\xi^{-1}(1-Z)]\\
&=&\int_{\delta/2}^{\delta+\epsilon(y)}[u(y)T'(1-x)-\lambda y F_\xi^{-1}(1-x)]dx
   -\int_\delta^{\delta+\epsilon(y)}[u(g(x))T'(1-x)-\lambda g(x)F_\xi^{-1}(1-x)]dx\\
&\ge&\int_{\delta/2}^{\delta+\epsilon(y)}[u(y)T'(1-x)-\lambda y F_\xi^{-1}(1-x)]dx
  -\int_\delta^{\delta+\epsilon(y)}u(g(x))T'(1-x)dx\\
&\ge& u(y)\int_{\delta/2}^\delta T'(1-x)dx -\lambda y\int_{\delta/2}^{\delta+\epsilon(y)}F_\xi^{-1}(1-x)dx\\
&=&y\left[\frac{u(y)}{y} \left(T(1-\delta/2)-T(1-\delta)\right)-\lambda \int_{\delta/2}^{\delta+\epsilon(y)}F_\xi^{-1}(1-x)dx\right].
\end{eqnarray*}
Since $\frac{u(y)}{y}\rightarrow +\infty$ as
$y\rightarrow 0$ and $T(1-\delta/2)-T(1-\delta)>0$, we have
$\frac{u(y)}{y} (T(1-\delta/2)-T(1-\delta))\rightarrow +\infty\;\;\mbox{ as } \;y\rightarrow 0.$ On the other hand,
$\epsilon(y)\rightarrow 0$ as $y\rightarrow 0$; hence
\[ \int_{\delta/2}^{\delta+\epsilon(y)}F_\xi^{-1}(1-x)dx\leq (\epsilon(y)+\delta/2)F^{-1}_\xi(1-\delta/2)\rightarrow \delta/2 F^{-1}_\xi(1-\delta/2)\;\;\mbox{ as } \;y\rightarrow 0.
\]
Consequently,
\[
\frac{u(y)}{y} (T(1-\delta/2)-T(1-\delta))-\lambda \int_{\delta/2}^{\delta+\epsilon(y)}F_\xi^{-1}(1-x)dx\rightarrow +\infty\;\;\mbox{ as } \;y\rightarrow 0.
\]
Fix $y>0$ sufficiently small so that the left hand side of the above
is no less than 1. Then
$$E[u(g_y(Z))T'(1-Z)-\lambda g_y(Z)F_\xi^{-1}(1-Z)]-E[u(g(Z))T'(1-Z)-\lambda g(Z)F_\xi^{-1}(1-Z)]\geq y>0,$$
which implies that $g_y$ is strictly better than $g$ for
(\ref{unc+part}). \eof

\begin{theo}\label{+optdis}
If $X^*$ is an optimal solution
for (\ref{plusutility})
with some $a>0$,
then $P(X^*=0)=0$.
\end{theo}
{\sc Proof:}
%
Proposition \ref{disopt} implies that
the distribution function $G^*$ of $X^*$ is optimal for (\ref{maxvg-a}).
By Lagrange method 
there exists $\lambda\geq0$ such that $(G^*)^{-1}$ is optimal for
(\ref{unc+part}). Since $a>0$, $(G^*)^{-1}\not\equiv 0$.
It follows then from Lemma \ref{Ginv>0} that $(G^*)^{-1}(x)>0$ $\forall x>0$,
or $G^*(0)=0$. \eof

So an optimal solution to (\ref{plusutility}) with a positive initial budget
is positive almost surely.

\medskip

\section{A Choquet Minimization Problem}\label{nppsubsection}

Consider a {general} utility minimization problem
involving the Choquet integral:
\begin{equation}\label{minusutility}
\begin{array}{ll}
\mbox{\rm Minimize}& V_2(X):=\int_0^{+\infty} T(P\{u(X)>y\})dy\\
\mbox{\rm subject to} & E[\xi X]=a, \;\; X\ge 0,
\end{array}
\end{equation}
where $\xi,\;a, \; T(\cdot)$ satisfy the same assumptions as those with
Problem (\ref{plusutility}), and $u(\cdot)$ is strictly increasing, concave
with $u(0)=0$.

It is easy to see that (\ref{minusutility}) always admits feasible solutions
(e.g., $X=x\id_{\xi\le \xi_0}$ is feasible with appropriate
$x\in \R, \xi_0\in \R$);
hence the optimal value of (\ref{minusutility})
is a finite nonnegative number.

In view of Theorem \ref{comono}, a similar argument to that in Appendix \ref{pppsubsection}
reveals that
the optimal solution $X^*$ to (\ref{minusutility})
must be in the form of $G^{-1}(F_\xi(\xi))$
for some distribution function $G(\cdot)$,
which can be determined by the following problem
\begin{equation}\label{minvg-a0}
  \begin{array}{ll}
 \mbox{\rm Minimize}& v_2(G):= \int_0^{+\infty}T(P\{u(G^{-1}(Z))>y\})dy\\
  \mbox{\rm subject to} & \left\{\begin{array}{l}
E[G^{-1}(Z)F_\xi^{-1}(Z)]=a,\\
G \mbox{ is the distribution function of a nonnegative random variable},
\end{array}\right.
  \end{array}
\end{equation}
where $Z:=F_\xi(\xi)$.

\begin{prop}\label{disopt2}
  If $G^*$ is optimal for (\ref{minvg-a0}), then
$X^*:=(G^*)^{-1}(Z)$
  is optimal for (\ref{minusutility}).
Conversely, if $X^*$ is optimal for
(\ref{minusutility}), then
its distribution function $G^*$ is optimal for
(\ref{minvg-a0}) and $X^*=(G^*)^{-1}(Z),\;\as$.
\end{prop}

By the same calculation as in Appendix \ref{pppsubsection}, we have
$v_2(G)=E[u(G^{-1}(Z)) T'(1-Z)]$. Denoting $g=G^{-1}$,
Problem (\ref{minvg-a0}) can be rewritten as
\begin{equation}\label{minvg-a}
  \begin{array}{ll}
\mbox{\rm Minimize}& \bar v_2(g):=E[u(g(Z)) T'(1-Z)]\\
\mbox{\rm subject to} & E[g(Z)F_\xi^{-1}(Z)]=a,\;\;
 g\in \Gamma.
  \end{array}
\end{equation}

Since the objective of the above problem is to {\it minimize}
a {\it concave} functional,  its solution must have a very
different structure compared with Problem (\ref{plusutility}), which
in turn requires a completely different technique
to obtain. Specifically, the solution should be a ``corner point solution''
(in the terminology of linear program). The question is
how to characterize
such a corner point solution in the present setting.

\begin{prop}\label{minv}
Assume that $u(\cdot)$ is strictly concave at $0$. 
Then the optimal solution for Problem (\ref{minvg-a}), if it exists,
must be in the form $g(t)=q(b)\id_{(b,1)}(t)$, $t\in[0,1)$,
with some $b\in [0,1)$ and $q(b):=\frac{a}{E[F_\xi^{-1}(Z)\id_{(b,1)(Z)}]}$.
\end{prop}
{\sc Proof:}
Denote $f(\cdot):=F_\xi^{-1}(\cdot)$ for notational convenience.
We assume $a>0$ (otherwise the result holds trivially).
If $g$ is an optimal solution to (\ref{minvg-a}),
then $g\not\equiv 0$.
Fix $t_1\in (0,1)$ such that $g(t_1)>0$. Define
$k:=\frac{\int_0^1g(t)f(t)dt}{\int_0^{t_1}g(t)f(t)dt+g(t_1)\int_{t_1}^1f(t)dt}$, and
$$\bar g(t):=\left\{\begin{array}{lll}
                  kg(t),&& \mbox{ if }t\in [0,t_1]\\
                  kg(t_1)&&\mbox{ if }t\in (t_1,1).\\
                \end{array}\right.
$$
Then $\bar g(\cdot)\in \Gamma$, and
$\int_0^1\bar g(t)f(t)dt=k\int_0^{t_1}g(t)f(t)dt+kg(t_1)\int_{t_1}^1f(t)dt=\int_0^1g(t)f(t)dt$,
implying that $\bar g(\cdot)$ is feasible for (\ref{minvg-a}).
We now claim that $g(t)=g(t_1),\;\; \aee t\in(t_1,1)$.
Indeed,
if this is not true, then $k>1$. Define $\lambda:=1-1/k\in (0,1)$ and
$\tilde g (t):=\frac{g(t)-g(t_1)}{\lambda}\id_{t>t_1},\;t\in[0,1)$.
Then
\begin{equation}\label{concombi}
(1-\lambda)\bar g(t)+\lambda \tilde g (t)=g(t)\;\;\forall t\in[0,1).
\end{equation}
It follows from the concavity of $u(\cdot)$ that
$\bar v_2(g)\ge (1-\lambda)\bar v_2(\bar g)+\lambda \bar v_2(\tilde g )$, and the equality holds only if
$$u(g(t))=(1-\lambda)u(\bar g(t))+\lambda u(\tilde g (t)),\;\; \aee t\in (0,1).$$
Owing to the optimality of  $g$, the above equality does hold.
However, the equality when $t\le t_1$ implies that
$u(\cdot)$ is {\it not} strictly concave at $0$, which is a contradiction.

Denote $b:=\inf\{t\ge0: g(t)>0\}$. The preceding analysis shows that
$g(t)=k\id_{t>b}$ for some $k\in \R^+$. The feasibility of $g(\cdot)$
determines
$k\equiv q(b)=\frac{a}{E[F_\xi^{-1}(Z)\id_{(b,1)(Z)}]}$.
\eof

{\rm By left-continuity one can extend the optimal $g$ described
in Proposition \ref{minv} to $[0,1]$ by defining $g(1):=q(b)$.
Moreover, since $g(t)$ is uniformly bounded in $t\in[0,1]$, it follows from
Proposition \ref{disopt2} that any optimal solution $X^*$ to
(\ref{minusutility}) can be represented as
$X^*=g(Z)$, hence must be uniformly bounded from above. }

Proposition \ref{minv}
suggests that we only need to find an optimal {\it number} $b\in[0,1)$
so as to solve Problem (\ref{minvg-a}), which motivates the
introduction of the following problem
\begin{equation}\label{minvc-a}
  \begin{array}{ll}
\mbox{\rm Minimize}& \tilde v_2(b):= E[u(g(Z)) T'(1-Z)]\\
\mbox{\rm subject to} & g(\cdot)=\frac{a}{E[F^{-1}_{\xi}(Z)\id_{(b,1]}(Z)]}\id_{(b,1]}(\cdot), \;\;0\le b<1.
  \end{array}
\end{equation}

\begin{prop}\label{VC=VG}
 Problems (\ref{minvg-a}) and (\ref{minvc-a})
have the same infimum values.
\end{prop}

{\sc Proof:}
Denote by $\alpha$ and $\beta$ the infimum values of Problems (\ref{minvg-a}) and (\ref{minvc-a})
respectively. Clearly $\alpha\le \beta $. If the opposite
inequality is false, then
there is  a feasible solution $g$ for (\ref{minvg-a})
such that $\bar v_2(g)<\beta $.

For any $s\in [0,1)$, define $k(s)=\frac{\int_0^1g(t)f(t)dt}{\int_0^sg(t)f(t)dt+g(s)\int_s^1f(t)dt}\ge 1$,
where $f(\cdot):=F^{-1}_\xi(\cdot)$.
Then $\lim_{s\rightarrow 1}k(s)=1$. Define
$$ H_s(t):=\left\{\begin{array}{lll}
                  k(s)g(t),&& t\in [0,s]\\
                  k(s)g(s),&&t\in (s,1).\\
                \end{array}\right.
$$
As shown in the proof of Proposition \ref{minv},
$H_s(\cdot)$ is feasible for ({\ref{minvg-a}), and
\begin{eqnarray*}
\bar v_2(H_s)&\le& \int_0^1u(k(s)g(t))T'(1-t)dt\\
&\le& \int_0^1k(s)u(g(t))T'(1-t)dt\rightarrow \bar v_2(g), \;\;\mbox{as }
s\rightarrow 1.
\end{eqnarray*}
Therefore
there exists $s\in [0,1)$, which we now fix,
such that $v_2(H_s)<\beta $.
For any nonnegative integer $n$, define $a(n,k):=\frac{\int_{(k-1)/2^n}^{k/2^n}H_s(t)f(t)dt}{\int_{(k-1)/2^n}^{k/2^n}f(t)dt}$,
for any $k=1,\cdots, 2^n$. It is clear
that $H_s((k-1)/2^n)\le a(n,k)\le H_s(k/2^n)$. Define
$$g_n(t):=\sum_{k=1}^{2^n}a(n,k)\id_{((k-1)/2^n, k/2^n]}(t),\;\;t\in[0,1).$$
Clearly $g_n\in\Gamma$, and $\int_0^1g_n(t)f(t)dt=\int_0^1H_s(t)f(t)dt=a$,
implying that $g_n$ is feasible for (\ref{minvg-a})
for each $n$.
Furthermore, $g_n(t)\rightarrow H_s(t)\;\;\forall t$ and
$0\leq g_n(t)\le k(s)g(s) \;\;\forall t$, which leads to
$\bar v_2(g_n)\rightarrow \bar v_2(H_s)$.  So there exists $n$ such that $\bar v_2(g_n)<\beta $.

Because $g_n(\cdot)$ is a left continuous and non-decreasing step function, we can
rewrite  it as
$$g_n(t)=\sum_{k=1}^m a_{k-1}\id_{(t_{k-1},t_k]}(t)$$
with $0=t_0<t_1<\cdots<t_m=1$, $0=a_0<a_1<a_2<\cdots<a_m<+\infty$.
Denote $\lambda_k:=\frac{a_k-a_{k-1}}{q(t_k)}$, $k=1,2,\cdots,m$,
where $q(\cdot)$ is defined in Proposition \ref{minv}.
Then for any $t\in (0,1)$,
\[ g_n(t)=\sum_{k=1}^m a_{k-1}\id_{(t_{k-1},t_k]}(t)
=\sum_{k=1}^m (a_k-a_{k-1})\id_{(t_k,1]}=\sum_{k=1}^m\lambda_kJ_{t_k}(t),
\]
where $J_{t_k}(t):=q(t_k)\id_{(t_k,1]}$.
Since
\[
a\equiv  \int_0^1g_n(t) f(t)dt=\sum_{k=1}^m\lambda_k\int_0^1J_{t_k}(t)f(t)dt=\sum_{k=1}^m\lambda_k a,
\]
we conclude that $\sum_{k=1}^m\lambda_k=1$, which means that
$g_n$ is a convex combination of $J_{t_k}$.
It follows from the concavity of $u(\cdot)$
that there exists $k$ such that $\bar v_2(J_{t_k})\le \bar v_2(g_n)$,
which contradicts the conclusion that $\bar v_2(g_n)<\beta \le \bar v_2(J_{t_k})$.
\eof

Summarizing, we have the following result.
\begin{theo}\label{-general}
Problems (\ref{minusutility}) and (\ref{minvc-a})
have the same infimum values.
If, in addition, $u(\cdot)$ is strictly concave at $0$, then
  (\ref{minusutility}) admits an optimal solution if and only if
the following problem
  $$\min_{0\le c< {\rm esssup}\;\xi}u\left(\frac{a}{E[\xi\id_{\xi>c}]}\right)T(P(\xi>c))$$
  admits an optimal solution $c^*$, in which case
the optimal solution to (\ref{minusutility})
is $X^*=\frac{a}{E[\xi\id_{\xi>c^*}]}\id_{\xi>c^*}$.
\end{theo}
{\sc Proof:}
The first conclusion follows from Proposition \ref{VC=VG}.
For the second conclusion, we rewrite the objective functional
of (\ref{minvc-a}) as
\[
\tilde v_2(b)=E\left[u\left(q(b)\id_{(b,1]}(Z)\right)T'(1-Z)\right]
=\int_b^1u(q(b))T'(1-t)dt=u(q(b))T(1-b),
\]
where $b\in[0,1)$. Now let $c:=F_\xi^{-1}(b)\in[0,{\rm esssup}\;\xi)$.
Then
\[\tilde v_2(b)=u(q(b))T(1-b)=u\left(\frac{a}{E[\xi\id_{\xi>c}]}\right)T(P(\xi>c)),\]
 and the desired results are straightforward in view of Theorem \ref{minv}.
\eof

\medskip

\section{Replicating a Binary Option}\label{replicate}

In this subsection, we want to find a portfolio replicating
the contingent claim $\rho^\alpha\id_{\rho\in (c_1, c_2)}$,
where $0\le c_1<c_2\le +\infty$, $\alpha\in \R$, and $\rho=\rho(T)$ with
\[ \rho(t):=\exp\left\{-(r+\frac{1}{2}|\theta|^2)t-\theta'W(t)\right\},\;\;0\le t\le T.
\]
The claim resembles the payoff of a binary (or digital) option, except that
$\rho$ does not correspond to any underlying stock [although it is
indeed the terminal wealth of a mutual fund; see Bielecki {\it et al.} (2005,
Remark 7.3), for details].

Let $\psi(\cdot)$ and $N(\cdot)$ be
the density function and distribution function of the
standard normal distribution respectively.
Recall that $\rho(t,T):=\rho(T)/\rho(t)$ conditional on
${\cF}_t$ follows a lognormal distribution with
parameters $(\mu_t,\sigma_t^2)$ given by (\ref{log}).

\begin{theo}  \label{degital}
If $c_2<+\infty$, then the wealth-portfolio pair
replicating $\rho^\alpha\id_{\rho\in (c_1, c_2)}$ is
\[ \begin{array}{l}
x(t)=\frac{\rho(t)^\alpha}{\sigma_t} \int_{c_1/\rho(t)}^{c_2/\rho(t)}y^\alpha \psi \left(\frac{\ln y -\mu_t}{\sigma_t}\right)dy,\\
\\
\pi(t)=-\left[\alpha x(t)-\frac{1}{\sigma_t\rho(t)}\left(
  c_2^{\alpha+1}\psi\left(\frac{\ln c_2-\mu_t-\ln \rho(t)}{\sigma_t}\right)
  -c_1^{\alpha+1}\psi\left(\frac{\ln c_1-\mu_t-\ln \rho(t)}{\sigma_t}\right)
  \right)\right](\sigma\sigma')^{-1}B.
\end{array}
\]
If $c_2=+\infty$, then the corresponding replicating pair is
\[ \begin{array}{l}
 x(t)=\frac{\rho(t)^\alpha}{\sigma_t} \int_{c_1/\rho(t)}^{+\infty}y^\alpha
\psi\left(\frac{\ln y -\mu_t}{\sigma_t}\right)dy,\\
\\
  \pi(t)=-\left[\alpha x(t)+\frac{1}{\sigma_t\rho(t)} c_1^{\alpha+1}\psi\left(\frac{\ln c_1-\mu_t-\ln \rho(t)}{\sigma_t}
  \right)\right](\sigma\sigma')^{-1}B.
\end{array}
\]
\end{theo}

\newcommand{\pd}[2]{\frac{\partial #1}{\partial #2}}

\pf When $c_2<+\infty$, the replicating wealth process is
\begin{eqnarray*}
  x(t)&=&E[\rho(t,T)\rho(T)^\alpha\id_{\rho(T)\in (c_1, c_2)}|\cF_t]\\
  &=&\rho(t)^\alpha E[\rho(t,T)^{\alpha+1}\id_{\rho(t,T)\in (c_1/\rho(t), c_2/\rho(t))}|\cF_t]\\
  &=&\rho(t)^\alpha \int_{c_1/\rho(t)}^{c_2/\rho(t)}y^{\alpha+1}dN\left(\frac{\ln y -\mu_t}{\sigma_t}\right)\\
  &=&\frac{\rho(t)^\alpha}{\sigma_t} \int_{c_1/\rho(t)}^{c_2/\rho(t)}y^\alpha
\psi\left(\frac{\ln y -\mu_t}{\sigma_t}\right)dy=f(t,\rho(t)),
\end{eqnarray*}
where $f(t,\rho):=\frac{\rho^\alpha}{\sigma_t} \int_{c_1/\rho}^{c_2/\rho}y^\alpha\psi\left(\frac{\ln y -\mu_t}{\sigma_t}\right)dy.$
It is well known that the replicating portfolio is
\begin{equation}\label{repi}
 \pi(t)=-(\sigma\sigma')^{-1}B\pd{f(t,\rho(t))}{\rho}\rho(t);
\end{equation}
see, e.g., Bielecki {\it et al.} (2005, Eq. (7.6)).
Now we calculate
\begin{eqnarray*}
  \pd{f(t,\rho)}{\rho}&=&\frac{\alpha \rho^{\alpha-1}}{\sigma_t}\int_{c_1/\rho}^{c_2/\rho}y^\alpha \psi\left(\frac{\ln y -\mu_t}{\sigma_t}\right)dy\\
&&\;  +\frac{\rho^\alpha}{\sigma_t}\left[\left(\frac{c_2}{\rho}\right)^\alpha \psi\left(\frac{\ln c_2-\mu_t-\ln \rho}{\sigma_t}\right)\frac{-c_2}{\rho^2}-
  \left(\frac{c_1}{\rho}\right)^\alpha \psi\left(\frac{\ln c_1-\mu_t-\ln \rho}{\sigma_t}\right)\frac{-c_1}{\rho^2}\right]\\
  &=&\frac{\alpha x(t)}{\rho}-\frac{1}{\sigma_t\rho^2}\left[c_2^{\alpha+1}\psi\left(\frac{\ln c_2-\mu_t-\ln \rho}{\sigma_t}\right)
  -c_1^{\alpha+1}\psi\left(\frac{\ln c_1-\mu_t-\ln \rho}{\sigma_t}\right)\right].
\end{eqnarray*}
Plugging in (\ref{repi}) we get the desired result.

The case with $c_2=+\infty$ can be dealt with similarly (in fact more easily).
\eof
}

\bigskip


\end{document}